\documentclass{aa}  
\usepackage{natbib}
\bibpunct{(}{)}{;}{a}{}{,} 
\usepackage{graphicx}
\usepackage{txfonts}
\usepackage{subfig}
\usepackage{soul}
\usepackage[dvipsnames]{xcolor}
\usepackage{amsmath}
\usepackage{hyperref}
\hypersetup{
    colorlinks=true,
    citecolor=blue,
    linkcolor=blue,
    filecolor=magenta,      
    urlcolor=cyan}
\usepackage{caption}
\usepackage{lipsum}
\usepackage{animate}
\usepackage[normalem]{ulem}

\newcommand{\luv}{L_{\rm UV}}
\newcommand{\lx}{L_{\rm X}}

\begin{document}

   \title{Quasars as Standard Candles}

   \subtitle{V. Evaluation of a $\leq$ 0.06 dex intrinsic dispersion in the  $L_{\rm X}$\,--\,$L_{\rm UV}$ relation }

  \author{Matilde Signorini\inst{1,2,3}\thanks{\email{matilde.signorini@unifi.it}}, Guido Risaliti\inst{1,2}, Elisabeta Lusso\inst{1,2}, Emanuele Nardini\inst{2}, Giada Bargiacchi\inst{4,5}, Andrea Sacchi\inst{6}, \and Bartolomeo Trefoloni\inst{1,2}}
  
   \institute{Dipartimento di Fisica e Astronomia, Universit\`a di Firenze, via G. Sansone 1, 50019 Sesto Fiorentino, Firenze, Italy
         \and
             INAF - Osservatorio Astrofisico di Arcetri, Largo Enrico Fermi 5, I-50125 Firenze, Italy
         \and
            University of California-Los Angeles, Department of Physics and Astronomy, PAB, 430 Portola Plaza, Box 951547, Los Angeles, CA 90095-1547, USA
         \and
             Scuola Superiore Meridionale, Largo S. Marcellino 10, 80138 Napoli, Italy
         \and
             Istituto Nazionale di Fisica Nucleare (INFN), Sez. di Napoli, Complesso Univ. Monte S. Angelo, Via Cinthia 9, 80126, Napoli, Italy
         \and
            Center for Astrophysics | Harvard \& Smithsonian, 60 Garden Street, Cambridge, MA 02138, USA
             }

\titlerunning{Quasars as Standard Candles V. Evaluation of the intrinsic dispersion}
\authorrunning{M. Signorini et al.}

   \date{\today}

  \abstract
{A characteristic feature of quasars is the observed non-linear relationship between their monochromatic luminosities at rest-frame 2500 Å and 2 keV. This relationship is evident across all redshifts and luminosities and, due to its non-linearity, can be implemented to estimate quasar distances and construct a Hubble Diagram for quasars.
Historically, a significant challenge in the cosmological application of this relation has been its high observed dispersion. Recent studies have demonstrated that this dispersion can be reduced by excluding biased objects from the sample. Nevertheless, the dispersion remains considerable ($\delta \sim 0.20$ dex), especially when compared to the Phillips relation for supernovae Ia. Given the absence of a comprehensive physical model for the relation, it remains unclear how much of the remaining dispersion is intrinsically tied to the relation itself and how much can be attributed to observational factors not addressed by the sample selection and by the choice of X-ray and UV indicators. Potential contributing factors include (i) the scatter produced by using X-ray photometric results instead of spectroscopic ones, (ii) the intrinsic variability of quasars, and (iii) the inclination of the accretion disc relative to our line of sight.
In this study, we thoroughly examine these three factors and quantify their individual contributions to the observed dispersion. Based on our findings, we argue that the intrinsic dispersion of the X-ray/UV luminosity relationship is likely below 0.06 dex. We also discuss why high-redshift subsamples can show a significantly lower dispersion than the average one.
}

\keywords{galaxies: active; quasars: general; quasars: supermassive black holes; methods: statistical}

\maketitle
%
\section{Introduction}
\label{intro}
Quasars, the most luminous and persistent sources in the Universe, are now observed up to redshift $z\simeq7.5$ \citep{Mortlock11, Banados18, Wang21}. Their spectral energy distribution (SED) spans from the radio to the X-ray band, with the most intense emission observed at optical--UV wavelengths \citep[e.g.][]{Sanders89,Richards06,Elvis2012}. The bulk of this emission is interpreted to originate in a optically thick accretion disc surrounding the central supermassive black hole (SMBH) \citep{Shakura73}. \\
A consistent feature of quasars is the presence of the non-linear relation between their X-ray and UV luminosities, often represented as $\log(L_{\rm X}) =\gamma \log(L_{\rm UV}) + \beta$, with a slope $\gamma\simeq0.6$ \citep[e.g.][]{Steffen06,Lusso10,Young2010}. This relation has been known for decades \citep[e.g.][]{Tananbaum79}, and it has been observed to hold over a wide range of redshift and luminosities. Its presence suggests a strong interaction between the UV-emitting accretion disc and the X-ray corona, but the exact mechanism behind it remains elusive. \\
The $\lx$--$\luv$ relation has far-reaching implications not only for quasar physics but also for cosmology. The non-linearity and consistency of the relation allow us to determine cosmological distances and therefore use quasars as standard candles. Such a possibility has been acknowledged since the discovery of the relation, but it had not been successfully implemented due to the high observed dispersion ($\sim$0.40 dex), which made any application for cosmological purposes challenging. However, recent studies have shown that most of this dispersion is not intrinsic, but arises from observational effects that can for the most part be removed by filtering out biased objects \citep{RL15, Lusso16, Lusso20}. By removing quasars affected by dust reddening, gas absorption, or that might suffer from Eddington bias, the dispersion reduces significantly, down to $\sim$0.20 dex, making quasars an actually useful cosmological tool. \\
This development has enabled the extension of the Hubble diagram to higher redshift values than those achievable with supernovae Ia (SN Ia). The extended Hubble diagram of quasars and supernovae abides by the predictions of a flat $\Lambda$CDM cosmology up to redshift $z\sim1.5$, but a 4$\sigma$ tension shows up at higher redshifts \citep{RL19_nature, Lusso20}. This tension has been confirmed when investigated in a cosmology-independent way \citep[e.g., ][]{bargiacchi21, bargiacchi22, Giambagli23}, and it has recently been 
retrieved also with a UV spectroscopically-validated sample of $\sim$1800 objects \citep{Signorini23_qascIV}.  \\
The remaining, rather high, dispersion of the relation between luminosities is still one of the biggest issues for both the implementation of quasars as standard candles and our understanding of the physical process behind it. Recent studies have started to address this question. For example, \cite{Sacchi22} reported a reduced dispersion of 0.09 dex for a subset of 30 quasars at redshift $3.0 < z < 3.3$. This subset, despite its specific selection criteria\footnote{This interval contains 15 sources for which high-quality XMM-\textit{Newton} pointed observations were obtained \citep{nardini19,Sacchi22}}, shows UV and X-ray properties consistent with the broader quasar population. Furthermore, \cite{Signorini23_qascIV} have demonstrated that, with the right selection of UV and X-ray proxies, it is possible to reduce the dispersion for a much larger sample from $\sim$0.20 dex down to $\sim$0.16 dex.\\
These findings suggest that the intrinsic dispersion of the $\lx$--$\luv$ relation is low and that much of the observed dispersion is linked to observational, and not intrinsic, factors. In this paper, we aim to investigate in detail these factors. We seek to quantify their contribution to the total observed dispersion and estimate the true intrinsic dispersion of the $\lx$--$\luv$ relation. This exploration is key to understanding the physical connection between the accretion disc and the corona and, by extension, the viability of quasars as cosmological probes. \\
In this work, we will address three main contributors to the dispersion: X-ray variability, the inclination of the accretion disc relative to our line of sight, and potential biases in X-ray flux estimates via photometry. We consider the total dispersion, $\delta_{\rm tot}$, as a combination of the intrinsic dispersion of the $\lx$--$\luv$ relation, $\delta_{\rm int}$, and the dispersion introduced by observational issues, $\delta_{\rm obs}$. With no universally accepted model explaining the $\lx$--$\luv$ relation yet, our approach focuses on determining $\delta_{\rm obs}$ to better constrain and understand $\delta_{\rm int}$. Through this approach, we aim to provide valuable insights into this relationship and support the implementation of quasars in cosmology. \\

The paper is structured as follows: in Section \ref{sec:sample} we describe the parent sample and the observational biases that have already been removed with its selection; in Section \ref{sec: variability} we investigate the contribution of the variability to the observed dispersion. In Section \ref{sec: inclination} we consider the contribution of inclination, providing an estimate through the use of a mock sample of quasars, while in Section  \ref{sec:xray} we focus on the possible contribution of using X-ray photometric data instead of spectroscopic ones. In Section \ref{sec: comparison} we compare our findings with the observational results about the dispersion, and in Section \ref{sec: conclusions} we draw our conclusions.

\section{Sample selection}
\label{sec:sample}
For this work, we considered the objects selected in  \citet[][hereafter L20]{Lusso20}. This sample is made of $\sim$2400 quasars, all of which have available UV and X-ray data in public catalogues. As discussed in L20, these objects have been selected in order to remove biased sources, and this process allowed us to reduce the observed dispersion from $\sim$0.40 dex to $\sim$0.20 dex. Here we briefly discuss these removable sources of bias; more details are given in Section 5 of L20. \\
\textbf{BAL and radio-loud quasars:} the UV fluxes of broad absorption line (BAL) quasars can be heavily affected by dust absorption and, as a consequence, these objects will deviate from the $\lx$--$\luv$ relation. Radio-loud quasars, instead, possibly show a jet-related X-ray component that adds to the one emitted by the corona, so they are supposed to deviate from the $\lx$--$\luv$ relation too. Quasars flagged as either BAL or radio-loud are therefore excluded from the sample. \\
\textbf{Dust extinction:} in principle, dust extinction could affect to some extent most of the observed quasars. Since dust reddening has a stronger impact on the UV band than on the X-rays, its contribution can significantly alter the $\lx$--$\luv$ relation. To minimize this effect, the sample was selected as described in \citet{Lusso16}: for each quasar the slopes $\Gamma_{1}$ and $\Gamma_{2}$ of the $\log(\nu) - \log(\nu L_{\nu})$ power-law continuum in the 0.3--1 $\mu$m and in the 1450--3000 {\AA} range were computed. Then, it was considered the point in the $\Gamma_{1}$--$\Gamma_{2}$ plane that corresponds to the SED of \cite{Richards06} with zero extinction, which is at $\Gamma_1 = 0.82$ and $\Gamma_2 = 0.40$.
The objects whose value of $\Gamma_{1}$--$\Gamma_{2}$ fell into the circle centred in 0.82--0.40 ($E(B-V)=0.0$), with a radius of 1.1 (which corresponds to $E(B-V)\simeq0.1$) were selected, while those which fell outside this range were removed from the sample. \\
\textbf{Eddington bias:} If the intrinsic X-ray flux of a quasar is close to the flux limit of a given observation, it will be detected only if a fluctuation towards higher fluxes takes place, while it will not be detected otherwise. This effect clearly introduces a potential bias in our sample. To avoid it, only sources that would remain above the flux limit even in case of a negative flux fluctuation were selected. \\
\textbf{X-ray absorption:} In the X-ray band, photoelectric absorption can take place due to the presence of gas in the quasar's rest frame. This gas absorption modifies the X-ray spectral shape; particularly, the effect in spectra on moderate quality is to lower the observed photon index $\Gamma$ of the X-ray continuum power law. The distribution of $\Gamma$ usually peaks around 1.9--2.0 for unobscured, luminous quasars \citep[e.g.,][]{Young09}, and $\Gamma$ exceeds 1.7 for almost all the sources with high-quality spectra. Therefore, to minimize this effect, only quasars with $\Gamma >1.7$ were selected. Furthermore, objects with $\Gamma>2.8$ are excluded too, to avoid soft-excess contribution at $z\la 1$ \cite[e.g.,][]{Sobolewska07, Gliozzi20}. 

\section{Variability}
\label{sec: variability}
Quasars exhibit non-periodic and stochastic variability across all observed wavelengths \citep{VandenBerk04, Markowitz04}, occurring on timescales ranging from hours \citep{Ponti12} to years \citep{deVries05, Vagnetti11}. Despite significant efforts, the underlying mechanisms driving this variability are still not completely understood. Notably, an observed anti-correlation between luminosity and variability has been established \citep{Hook94, Kelly09, Lanzuisi14, Paolillo17}, attributed to the influence of black-hole mass and accretion rate on the variability pattern.

\begin{figure}
	\centering
	\includegraphics[width=.99\linewidth]{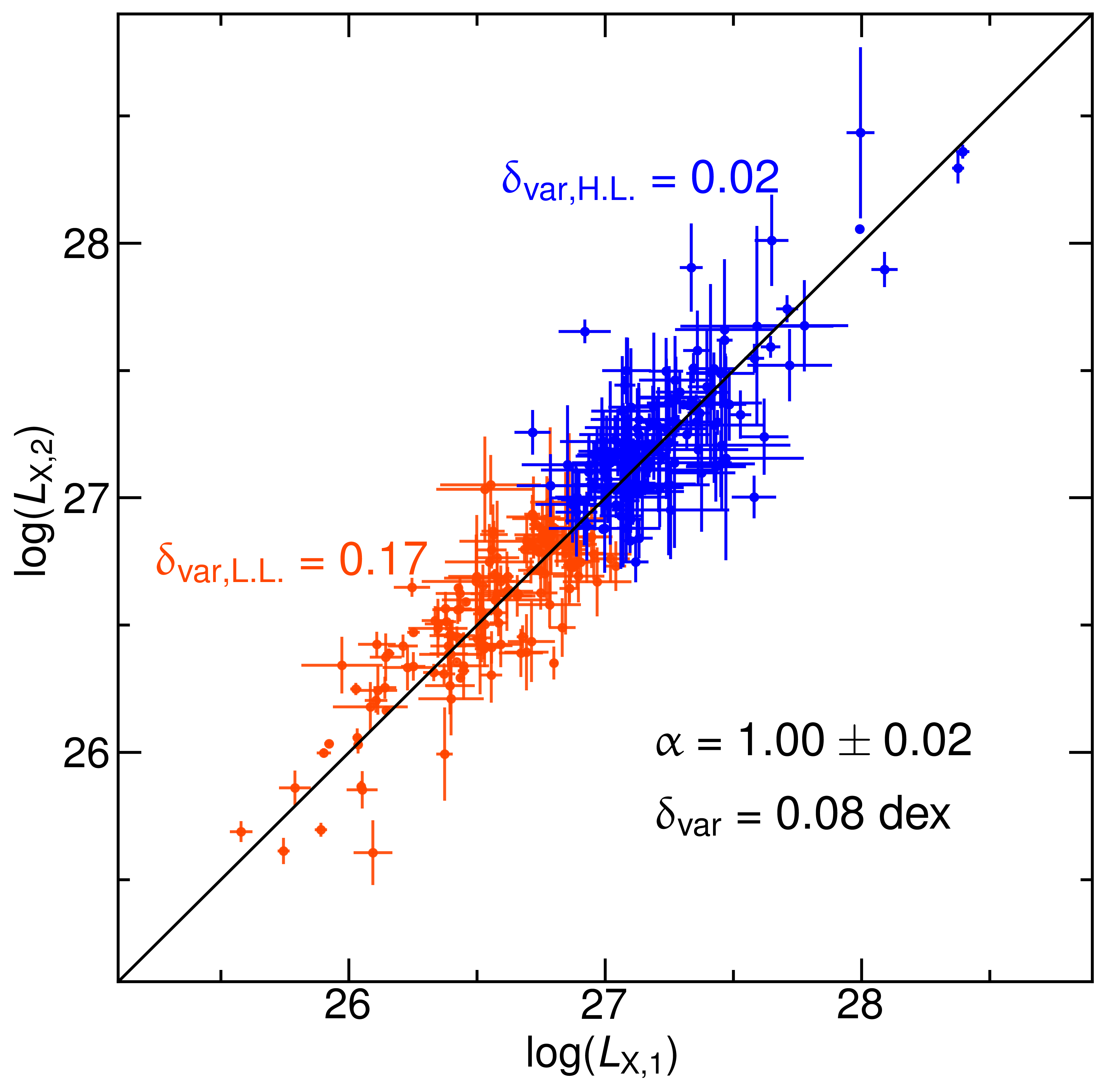}
	\caption{Comparison of the 2-keV luminosities for objects with multiple X-ray observations. The best fit is consistent with the bisector line (solid line). The dispersion parameter due to variability is $\delta_{\rm var}=0.08$ dex. When distinguishing between the high- and low-luminosity subsamples, we see that the dispersion parameter is higher for lower luminosities. Luminosities are derived from photometric fluxes, assuming a standard flat $\Lambda$CDM model. We note that, as we are comparing luminosities for the same object, the results do not depend on the chosen cosmological model }
	\label{fig:var_tot}
\end{figure}

The timescales of quasar variability are wavelength-dependent, with X-ray emission displaying considerably faster variability compared to the optical band. In the context of the $\lx$--$\luv$ relation this means that, because of the nature of variability, the same UV state corresponds to a range of X-ray states. 
Therefore, even if the intrinsic dispersion of the $\lx$--$\luv$ relation were null, this would introduce a scatter in the observed relation. Given that X-ray variability exhibits the largest amplitude on comparable timescales, it contributes most significantly to the observed dispersion. In this Section, we give an estimate of this contribution for the objects in the L20 sample. 

To do so, we looked for the objects in the L20 sample that have more than one serendipitous observation in the XMM-\textit{Newton} source catalogue 4XMM–DR9 \citep{Webb20}. We found 289 objects with multiple observations; the vast majority (80\%) has only two observations, so we considered the longest and the second longest observation for each object. Our goal was to compare the monochromatic 2 keV luminosities obtained from two observations that happened at random different times; we expected them to follow a one-to-one relation, with a scatter that would give us an estimate of the variability contribution to the dispersion in the $\lx$--$\luv$ relation. So we fit the relation between the second longest ($L_{\rm X,2}$) and the longest ($L_{\rm X,1}$) observations with a line: $\log(L_{\rm X,2}) = \alpha\,\log(L_{\rm X,1}) + \zeta$, with the slope $\alpha$ and the normalization $\zeta$ as free parameters. The fit was performed with a Bayesian approach of likelihood maximization; we used the \texttt{emcee} code, which is an implementation of Goodman $\&$ Weare’s Affine Invariant Markov Chain Monte Carlo (MCMC) Ensemble sampler \citep{emcee13}. In building the likelihood, we need to consider that we have uncertainties of similar magnitude on both axes; to account for this, we adapted the BCES method \citep{Akritas96}, where the tangent ellipse is used to measure the distance of each point form the best-fit line. 

The results are shown in Figure \ref{fig:var_tot}. We obtained, as expected, $\alpha = 1.00 \pm 0.05$ a slope, and an intercept value of $\zeta = -0.005 \pm 0.015$. We also derived the total dispersion of the relation as:
\begin{equation}
\delta_{\rm tot} = \sqrt{\sum_{\rm i}^{N} \log(L_{\rm X,2})^2_{i}-(\alpha \log(L_{\rm X,1})_{i}+\zeta)^2}    
\end{equation}
which turns out to be 0.17 dex. If the uncertainties on the $x$- and $y$-axis completely explained this dispersion, no intrinsic scatter due to variability would be present. However, we computed the average observational uncertainty on both $\log(L_{\rm X,1})$ and $\log(L_{\rm X,2})$ and we obtained 0.15 dex. Given this result, we can compute the intrinsic dispersion due to variability simply as $\delta_{\rm var} = \sqrt{\delta_{\rm tot}^2-\delta_{\rm err}^2}$, and get $\delta_{\rm var} = 0.08$ dex. This value can therefore be considered as the average contribution of X-ray variability to the dispersion in the $\lx$--$\luv$ relation. Although the number of objects for which we have multiple observations is only $\sim$12\% of the total L20 sample, these objects are fully representative (given that the observations we are considering are serendipitous) and they span the same luminosity range as the whole sample. So we can consider the result from this analysis to be applicable to the whole quasar catalogue. \\
We note that a similar analysis on objects with multiple observations was performed in \cite{Lusso16}, with a smaller sample of 159 objects, and they found the variability contribution to be $\delta_{\rm var} =0.12$ dex. Considering that the analysis described in this Section was performed with a sample which is more than doubled in statistics, we believe our estimate of $\delta_{\rm var} =0.08$ to be more accurate.\\
We also tested the dependence of variability on luminosity for our sample, by dividing it into two subsamples, above and below the value of $\log(L_{\rm X}) = 26.9$. The objects in the subsamples are 145 and 144, respectively, and for each of them we derived the total dispersion and the average uncertainty as described above. As can be seen in Fig. \ref{fig:var_tot}, the high-luminosity subsample shows a smaller dispersion due to variability ($\delta_{\rm var,H.L.}=0.02$ dex) than the low-luminosity one ($\delta_{\rm var,L.L.}=0.17$ dex). This is not only consistent with results in the literature, but it can, at least partially, also explain the results of \cite{Sacchi22}, where a dispersion as low as 0.09 dex is observed for a subsample of objects at redshift $3.0 < z < 3.3$. Together with having spectroscopic data and a subsample of pointed X-ray observations, these objects have an average luminosity of $\log(L_{\rm X}/{\rm erg\,s^{-1}\,Hz^{-1}}) \sim 27.7$, 
which is in the ``high luminosity'' regime. Therefore, the contribution of variability to their total dispersion must be very little, if not even zero. 

\section{Inclination}
\label{sec: inclination}
The second factor contributing to the observed dispersion we are considering, is the inclination of the quasar accretion disc relative to the line of sight. The optical--UV intrinsic emission from quasars is typically attributed to a disc-like component. The angle at which a quasar is viewed then crucially influences its observed flux. Specifically, unless the quasar is perfectly face-on, the observed flux, $f_{\rm obs}$, is derived from the intrinsic UV flux ($f_{\rm int}$) as $f_{\rm obs} = f_{\rm int} \, \cos\theta$, where $\theta$ is the angle between the observer's line of sight and the quasar disc axis. Notably, while the inclination affects the observed UV flux, it has no effect on the X-ray flux. This distinction arises because X-ray `coronal' emission is believed to be isotropic. The exact location and geometry of the corona are still largely unknown, although polarization results are now providing new perspectives on the topic \citep[e.g.,][]{Gianolli23}.  We note here that we will assume the X-ray emission to be isotropic throughout this work.\\
This inclination effect, by changing the UV flux, changes the slope of the relation, differently for each different inclination angle. Overall this results in an increase in the observed dispersion for the global quasar sample. 
Moreover, this effect is asymmetric, impacting quasars differently based on their relative brightness with respect to the detection limit. Bright quasars, surpassing by far the detection threshold, will still be detected even with a diminished observed flux $f_{\rm obs}$ at large inclination angles. However, for quasars nearing the flux limit, there is a range of $\cos\theta$ values where they become undetectable.\\
In an ideal scenario, if we had accurate knowledge of the inclination angle $\theta$ for every quasar observed, we could correct the observed flux to counterbalance inclination effects. Unfortunately, we do not have consistent estimators for quasars' inclination. Some works suggest that the intensity of the [O\,\textsc{iii}] line is an indicator of the inclination angle of a quasar \citep[e.g. ][]{Bisogni17}. Yet, this result is valid in a statistical sense, but can not be reliably used for an object by object correction. Furthermore, this line escapes the SDSS spectral range for redshifts above $\sim$0.7.\\
Given that we lack observational methods to measure inclinations directly, we will tackle this challenge using mock samples of quasars. While we might not be able to eliminate this source of dispersion, we can estimate its contribution. This is essential for accurately determining the intrinsic dispersion of the $\lx$--$\luv$ relation. In this Section, we discuss how to build a mock sample of quasars to correctly represent the effect of inclination, and derive an estimate for its contribution to the total dispersion.\\

\subsection{Correction to a luminosity function}

The first step to build our mock sample is to determine from which luminosity distribution we should extract our objects. Numerous studies have established the observed UV luminosity function for quasars and its redshift evolution \citep[e.g. ][]{Boyle00, Croom09, Ross13}. However, within a particular luminosity range, the quasars we observe are only those that, once inclined, have a flux above the survey flux limit. Consequently, the observed luminosity function for quasars is biased, and a priori we do not know how much the shape of the luminosity function changes because of that. Here we therefore attempt to deduce the intrinsic luminosity function of quasars from the observed one, and then use it as the starting point for our mock sample.\\
We designate $\theta$ as the inclination angle, signifying that a face-on quasar has $\theta=0$ whilst an edge-on quasar has $\theta = \pi/2$, as shown in Fig. \ref{fig: inc_sketch}. We postulate that quasars are randomly inclined in the sky. For a given intrinsic luminosity $\mathcal{L}$, the observed luminosity is $L = \mathcal{L} \cos\theta$, where $L = \mathcal{L}$ for face-on quasars and $L \rightarrow 0$ for those seen at increasingly large inclinations. 
Assuming the true quasar luminosity distribution in the Universe is a continuous function $m(\mathcal{L})$, and given that quasars are randomly inclined, 
a specific observed luminosity bin would contain objects with particular combinations of intrinsic luminosity and inclination angle. Hence, the observed distribution $n(L)$ can be expressed in terms of the intrinsic distribution $m(\mathcal{L})$:

\begin{equation}
	n(L)dL = \int_{0}^{\pi/2} m\left(\frac{L}{\cos\theta}\right)\sin\theta \,d\theta d\mathcal{L}
	\label{f1}
\end{equation}

Detailed calculations in Appendix \ref{appendix: LF} allow us to derive the true luminosity function $m(\mathcal{L})$ from the observed one, $n(L)$:

\begin{equation}
	m(\mathcal{L}) = n(L) - n'(L) L
	\label{eq_m}
\end{equation}

\begin{figure*}
	\centering
	\subfloat{{\includegraphics[width=8.75cm]{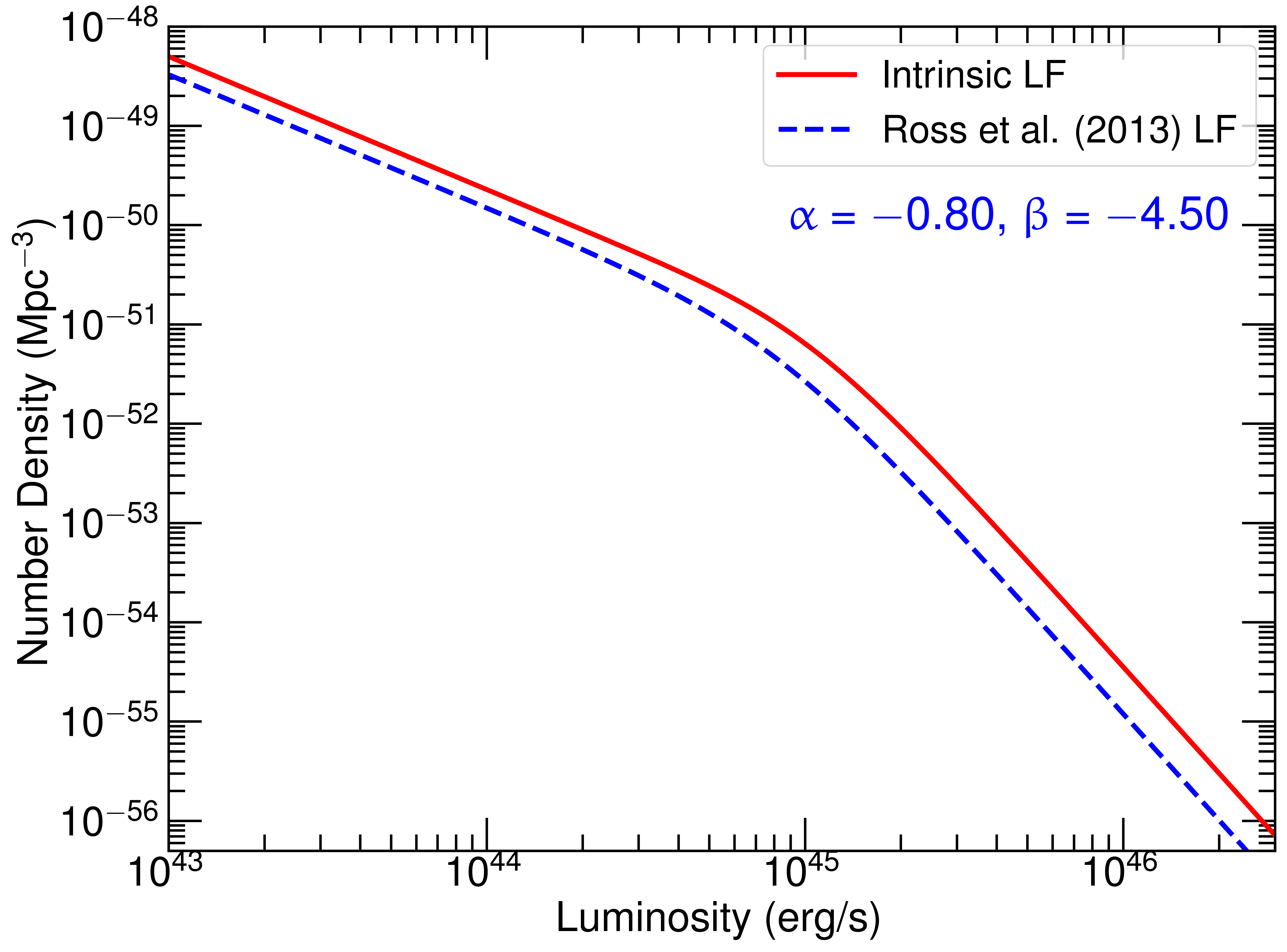} }}%
	\qquad
	\subfloat{{\includegraphics[width=8.75cm]{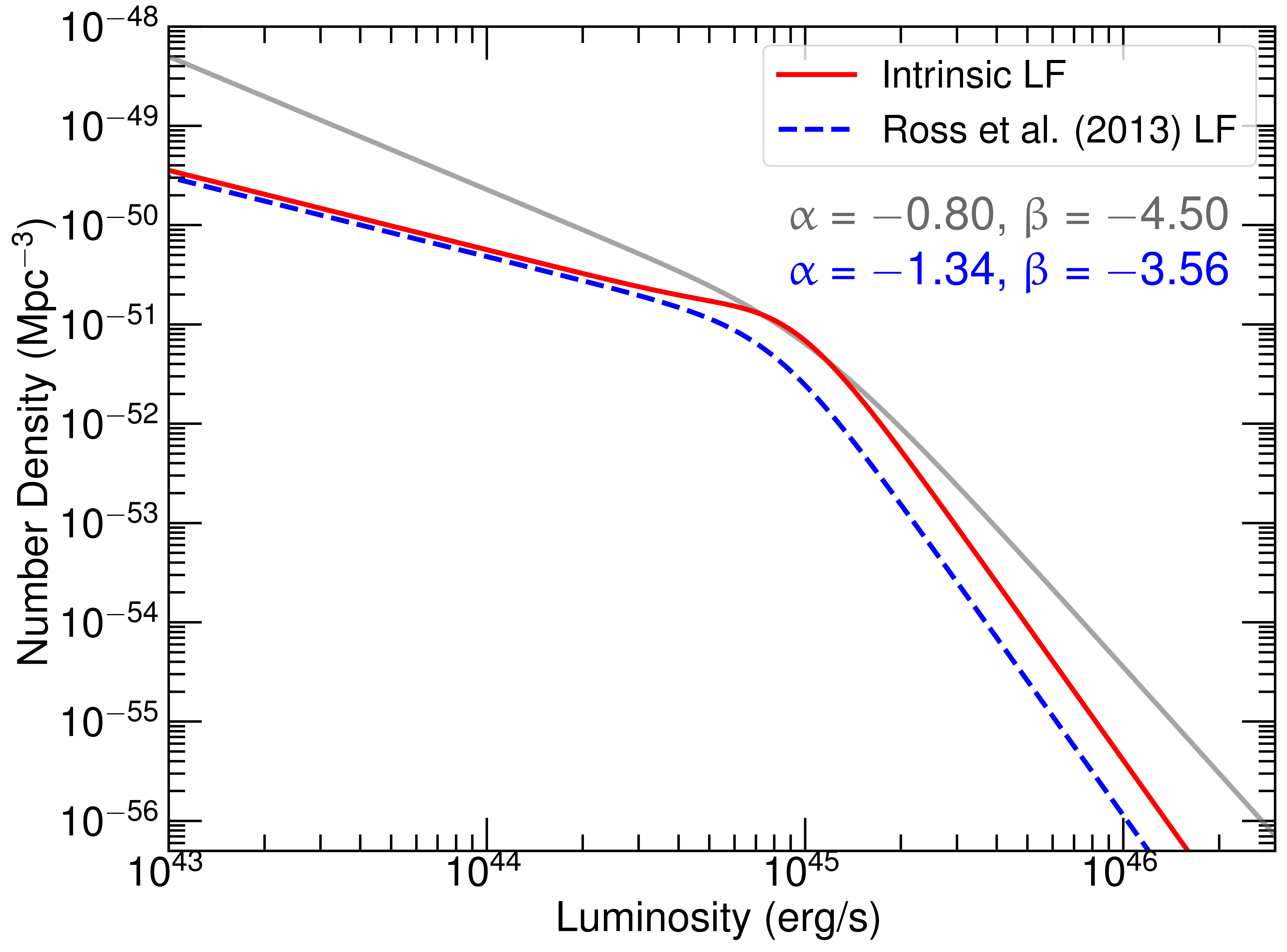} }}%
	\caption{\textit{Left}: observed luminosity function $n(L)$, in blue dashed, as per the \cite{Ross13} parametrization: the luminosity function is a broken power law with indexes $\alpha = -1.34$ and $\beta = -3.56$, and $\log(\phi_{*})$ = $-$6.15. A Pure Luminosity Evolution is assumed. In this Figure we show, as an example, the results for $z=1$. In solid red, the intrinsic luminosity function $m(\mathcal{L})$, which is corrected for inclination effects.
		\textit{Right}: same as the left panel, but with an observed luminosity function assumed to have indexes $\alpha =-0.8$ and $\beta = -4.5$. The comparison between the two panels shows that a greater difference between $\alpha$ and $\beta$ accentuates the `knee' distortion and the variance between $m(\mathcal{L})$ and $n(L)$ at higher luminosities. For a better comparison, the red solid line of the left panel is also plotted in the right panel, in grey.}
	\label{fig: croom09}
\end{figure*}

The result of this correction is therefore the distribution of intrinsic quasar luminosities, which would be the observed distribution if all the objects were face-on. To properly compare with the observed $n(L)$, we multiply $m(\mathcal{L})$ by the average $\cos\theta$, which is 0.5. In Figure \ref{fig: croom09}, we show, as an example, the juxtaposition the observed $n(L)$ with the derived intrinsic luminosity function $m(\mathcal{L})$, assuming the quasar luminosity function shape from \cite{Ross13}:

\begin{equation}
	\frac{dn}{dL} = \frac{\phi_{*}}{\left(\frac{L}{L^{*}}\right)^{-\alpha}+\left(\frac{L}{L^{*}}\right)^{-\beta}}
\end{equation}
with parameters $\alpha$ and $\beta$ again following the results from \cite{Ross13}: $\alpha = -1.34$ and $\beta = -3.56$, and $\log(\phi_{*}) = -6.15$. As in the \cite{Ross13} parametrization, we assume the quasar luminosity function to follow a Pure Luminosity Evolution (PLE). In Fig. \ref{fig: croom09}, as an example, we show the results assuming a redshift value of $z=1$. 
Notably, at both low and high luminosities, the shape of the intrinsic luminosity function aligns with the observed one. The main differences are found around the change in the slope, commonly called the `knee'. 
Altering the $\alpha$ and $\beta$ values impacts the relative shapes: a greater difference between $\alpha$ and $\beta$ accentuates the `knee' distortion and the variance between $m(\mathcal{L})$ and $n(L)$ at higher luminosities. The right panel of Fig. \ref{fig: croom09} showcases the results for $\alpha =-0.8$ and $\beta = -4.5$.

\subsection{Mock sample} \label{subsec: mock}
Now that we know how to derive an inclination-corrected luminosity function, we can use it to build our mock sample of quasars to determine the effect of inclination on the $\lx$--$\luv$ observed dispersion. We do so, at first, allowing the inclination angle to vary between 0 and $\pi/2$. 

\begin{figure*}
	\centering
	\subfloat{{\includegraphics[width=8.75cm]{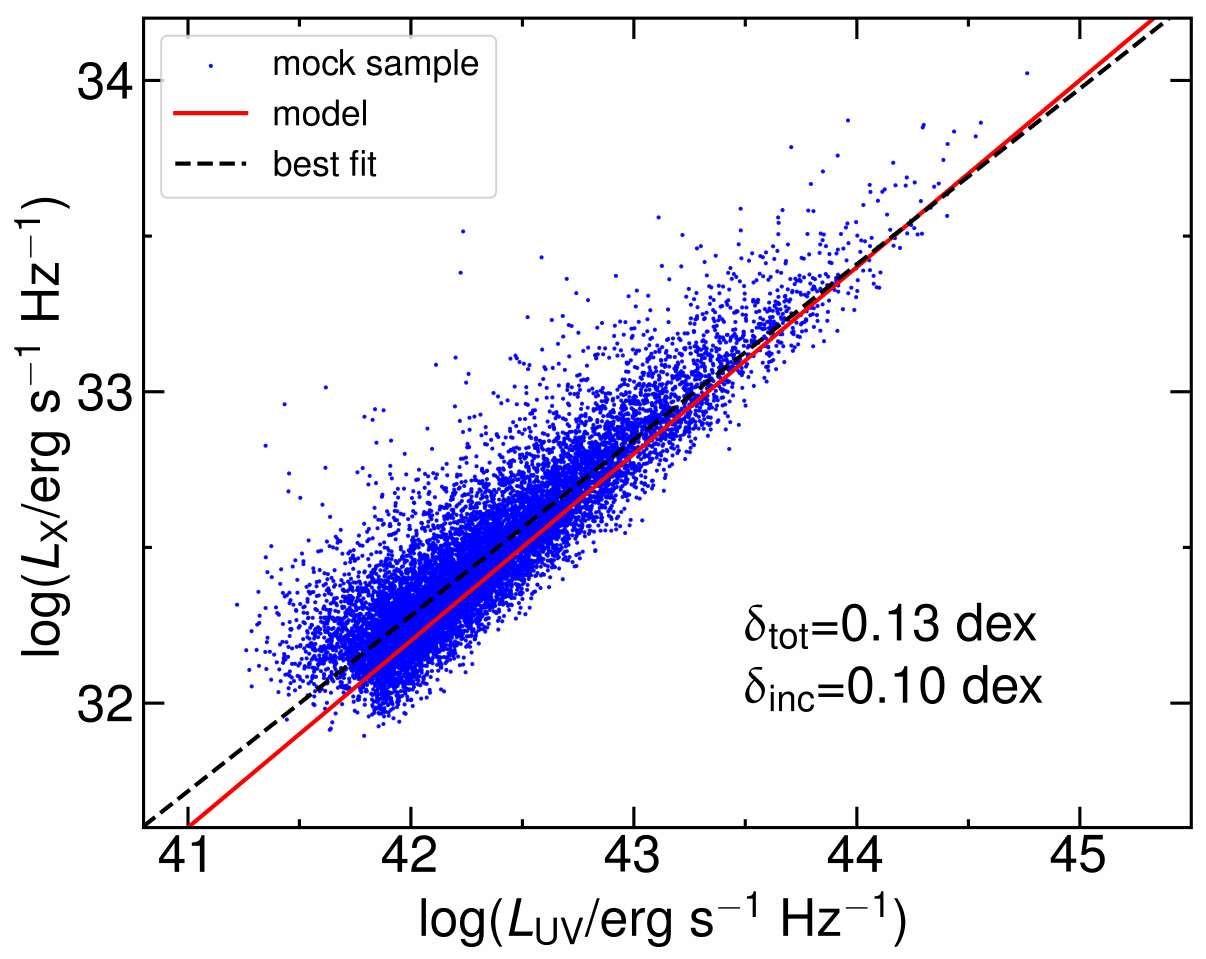} }}%
	\qquad
	\subfloat{{\includegraphics[width=8.75cm]{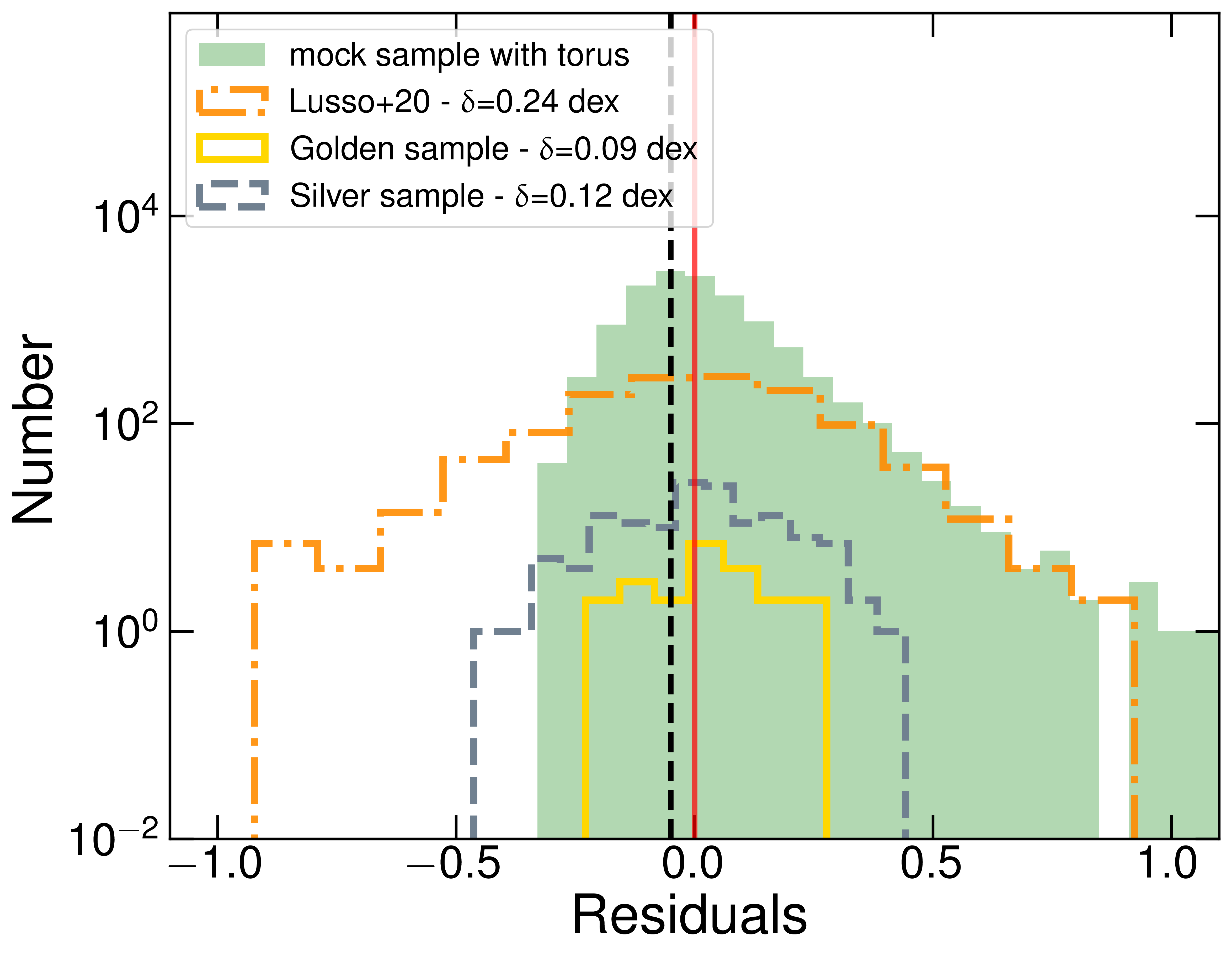} }}%
	\qquad
	\subfloat{{\includegraphics[width=8.75cm]{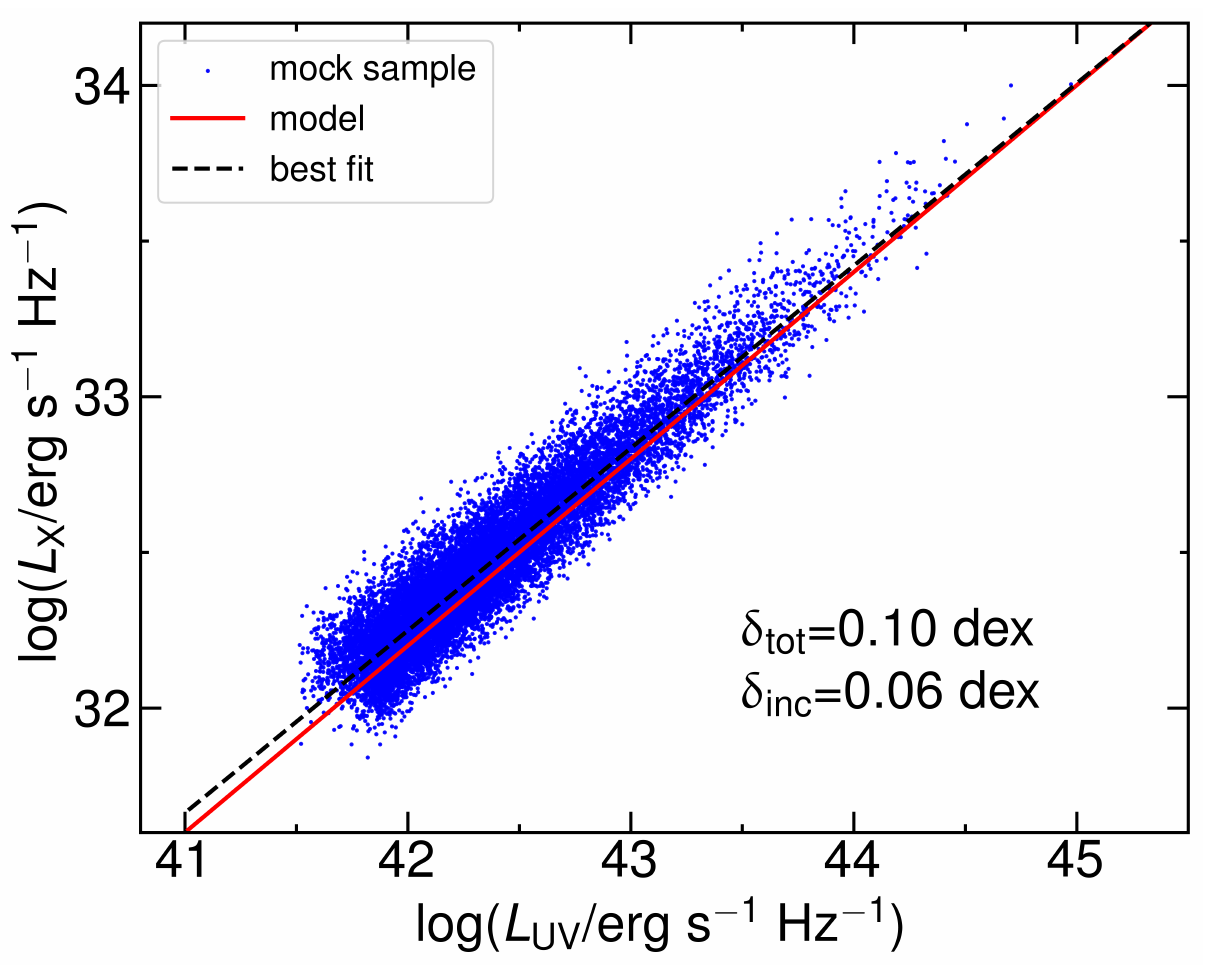} }}%
	\qquad
	\subfloat{{\includegraphics[width=8.75cm]{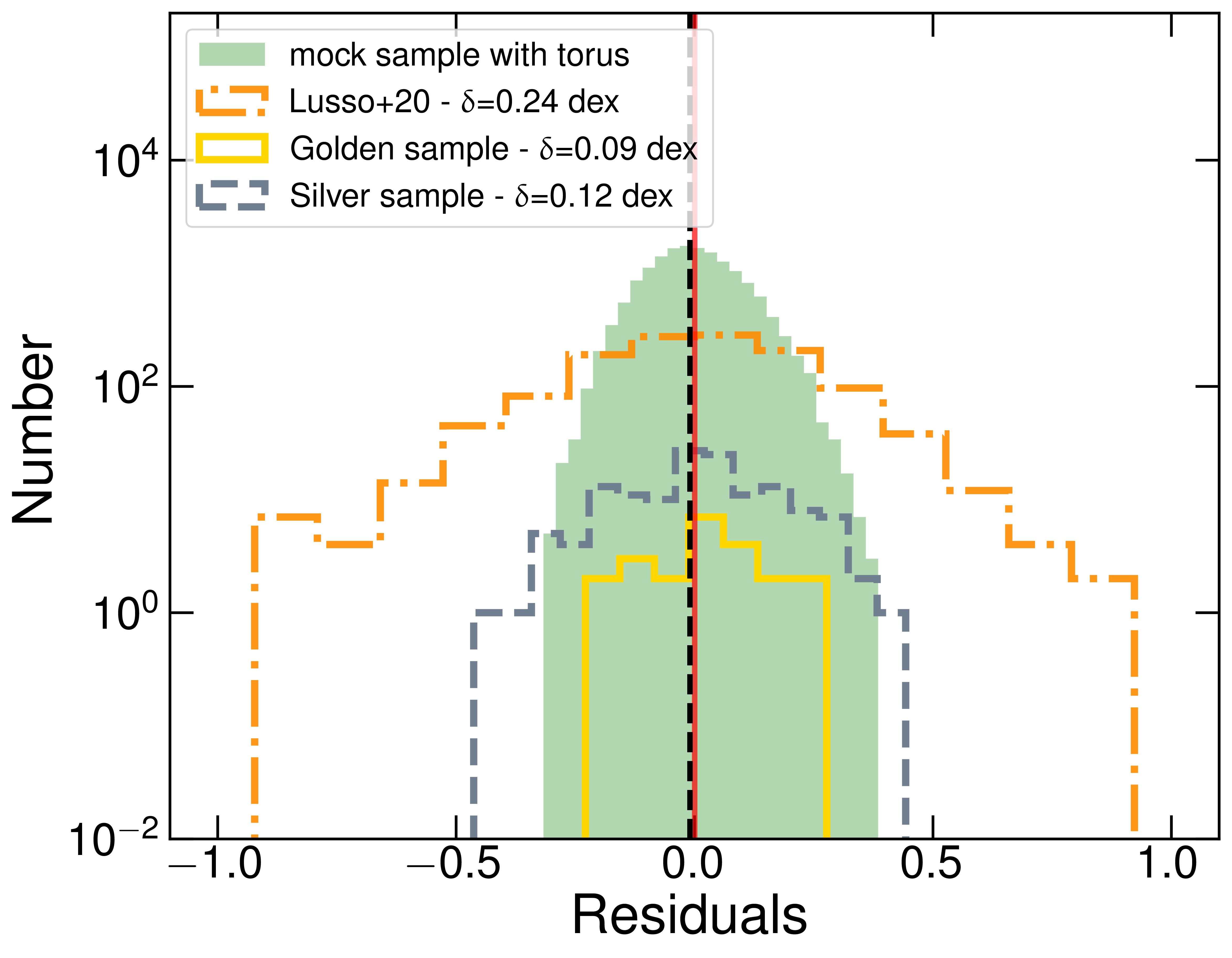} }}%
	\caption{
	\textit{Upper left panel}: mock sample of 100.000 quasars assuming an intrinsic relation with zero intrinsic dispersion, a contribution from variability to the observed dispersion of 0.08 dex, and an inclination angle between 0 and 90°. For each object, a redshift is assigned using the L20 sample redshift distribution; given the redshift, the corresponding luminosity function from \cite{Ross13} is derived, and the luminosity of the objects is extracted from there. The red solid line represents the starting sample, with a slope $\gamma=0.6$ and a zero dispersion. The blue points show the sample after the dispersion due to variability is added and the objects are assigned a random inclination. The total dispersion of the sample is $\delta_{\rm tot}=0.13$ dex, the inclination effect accounts for $\delta_{\rm inc}$=0.10 dex. \\
	\textit{Upper right panel}: in green filled, the histogram of the fit residuals. This distribution is skewed, with a skewness parameter of $s=1.18$. In dot-dashed orange, the histogram of the residuals for the L20 sample. We see that the L20 distribution, which is the observed one, is instead much more symmetric, with a skewness parameter of $s_{\rm L20} = 0.20$. In dashed silver and in solid gold, the residuals corresponding to the ``silver sample'' and the ``golden sample'' of \cite{Sacchi22} (details in Section \ref{sec: comparison}). The red solid vertical line corresponds to zero, while the dashed black vertical line corresponds to the peak of the mock sample distribution, equal to $-0.05$.  All the histograms are shown in logarithmic units to enhance the readability.\\  
	\textit{Lower left panel}: the same as the Upper Left panel, but assuming the presence of an obscurer with an angle width of $\psi_{\rm torus}=25\degr$, which means that the inclination angle for the objects in the mock sample can go from 0 to 65$\degr$.  The total dispersion is reduced to $\delta_{\rm tot}=0.10$ dex, and the inclination effect accounts for $\delta_{\rm inc}$=0.06 dex. \\
	\textit{Lower right panel}: the same as the Upper right panel, but assuming the presence of an obscurer with an angle width of $\psi_{\rm torus}=25\degr$. Here we see that the residuals distribution is symmetric, with a skewness parameter of $s=0.19$ which is consistent with the L20 value. 
}
	\label{fig: mock_ross_all}%
\end{figure*}

We start building a sample of 100,000 objects, and we consider the quasar luminosity function obtained by \cite{Ross13} for the SDSS as the starting point. Such a luminosity function assumes a redshift dependence. Therefore, we first assign a random redshift to the objects in the mock sample, extracting them from the redshift distribution of L20. Then, for each object, we derive the luminosity function corresponding to that redshift from \cite{Ross13}, we correct it for the inclination effect described in the previous subsection and we use it to extract a luminosity value for that object. The luminosity function described in \cite{Ross13} is derived for the $i$-magnitude, but we are interested in the monochromatic luminosity at 2500 {\AA}. Therefore, for each object, we derive it by assuming an SED with $f_{\rm \nu}\sim\nu^{-\alpha}$, with $\alpha=0.5$. We tested all of the following also assuming $\alpha=1$ and we always obtained consistent results. Given the 2500 {\AA} luminosities, we assume the $\lx$--$\luv$ relation with $\gamma$=0.6 and derive the corresponding $\log(L_{\rm X})$ for each object. The values of $\log(L_{\rm X})$ are then shifted by a random quantity extracted from a Gaussian with mean zero and standard deviation equal to 0.08; this mimics the contribution to the dispersion due to the variability, which, as we found out in Section \ref{sec: variability}, is $\delta_{\rm var}=0.08$. Then, for each object, we assign a random value of the inclination angle $\theta$, from 0 to $\pi/2$, as we are here assuming no absorbing torus. The angle is extracted from a distribution which is uniform in $\cos\theta$. 
Given $\theta$, the inclined UV luminosity is derived as $L_{\rm obs} = \mathcal{L} \cos\theta$, while the X-ray luminosity is left untouched. We now have a mock sample of inclined objects and we have to consider the presence of an observational threshold. We consider as a threshold the $i$-magnitude flux limit of the first SDSS release \citep{Richards02}, $m_i = 19.1$, from which we derived the corresponding monochromatic flux at 2500 {\AA} \footnote{The flux selection for the true observed sample is going to be much more complicated than a simple flux threshold. However, here we want to recreate a simpler scenario of a uniform sample with a given flux limit. We tested that the final results in terms of the inclination contribution to the dispersion do not depend on the exact choice of the flux limit. }. So for each object, given its redshift, we derived its 2500 {\AA} flux by assuming a standard flat $\Lambda$CDM cosmology and then removed the object from the mock sample if it fell below the threshold. We now fit the relation between $\log(L_{\rm X})$ and $\log(L_{\rm UV})$, where $L_{\rm UV}$ has the inclination-affected values. The fit is done with a Bayesian approach of likelihood maximisation, assuming the function $\log(L_{\rm X}) = \gamma \log{L_{\rm UV}} + \beta$, where $\gamma$ and $\beta$ are two free parameters, and where we take the presence of a dispersion $\delta$ into account by modifying the likelihood function accordingly. As before, we performed the fit using the \textit{emcee} code. The mock sample and the results of the fit are shown in the top left panel of Fig. \ref{fig: mock_ross_all}. We retrieve a slope coefficient $\gamma_{\rm fit}=0.58\pm0.02$, consistent with the assumed value of $\gamma=0.6$. This reassures us that the inclination effect is not significantly affecting the slope of the relation. The total dispersion is $\delta_{\rm tot} = 0.13$ dex. Given that we assumed a dispersion of $\delta_{\rm var}= 0.08$ dex due to the variability, to obtain the contribution of the inclination we have to quadratically subtract the two, so that $\delta_{\rm inc} = \sqrt{\delta_{\rm tot}^{2}-\delta_{\rm var}^2}$. The result is $\delta_{\rm inc} = 0.10$ dex. On the top right panel of Fig. \ref{fig: mock_ross_all} we show, in green, the histogram of the fit residuals (which we obtain by simply subtracting the best-fit model from the mock data). In dot-dashed orange, we show the same histogram for the L20 sample. It is clear that the point distribution of our mock sample is not representative of the observed scenario, as it is (slightly) off-centred and significantly skewed, with a skewness parameter of $s=1.18$, while the L20 distribution is much more symmetric, with a skewness parameter of $s_{\rm L20} = 0.2$. In the top right panel, we also show the distribution of the residuals for the ``silver'' and the ``golden'' samples of \cite{Sacchi22} which will be further discussed in Section \ref{sec: comparison}.

\begin{figure}
	\centering
	\includegraphics[width=.99\linewidth]{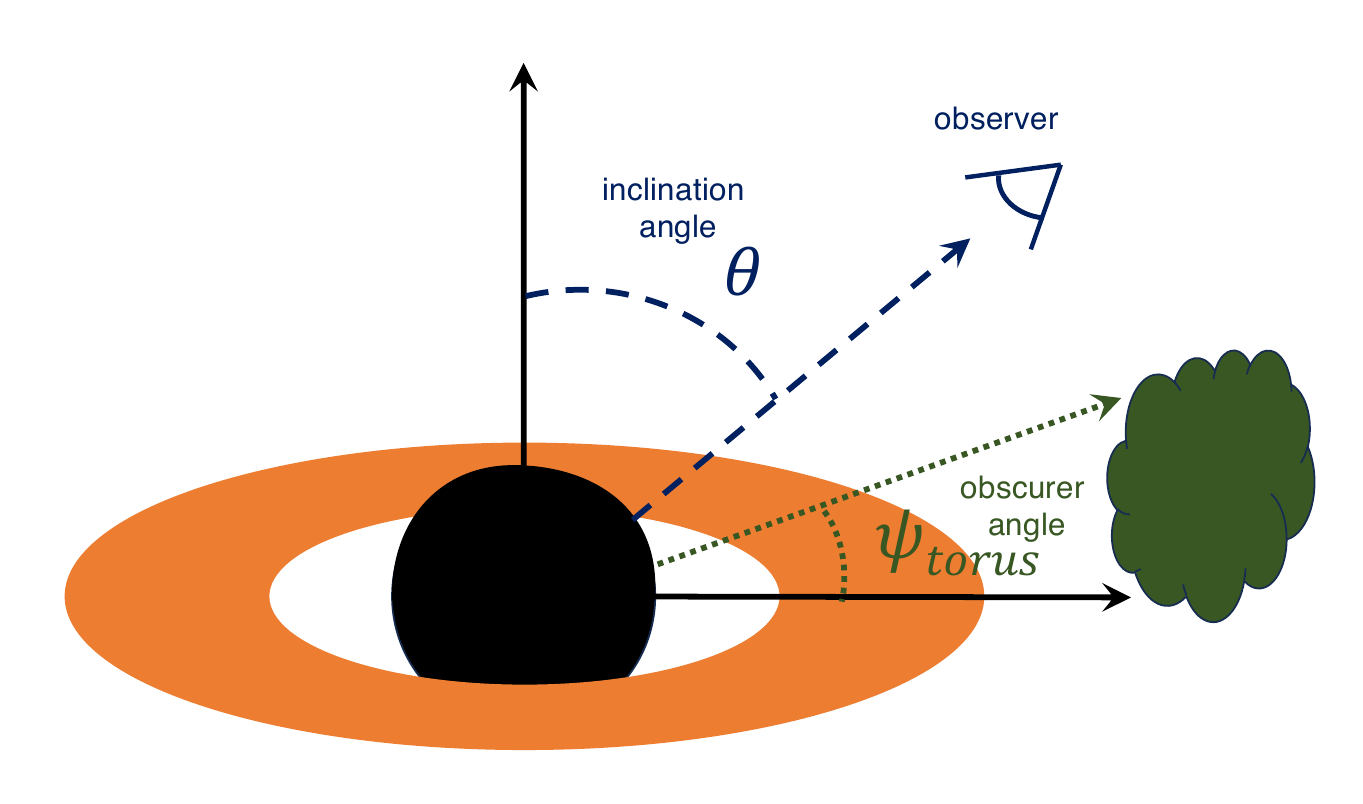}
	\caption{Schematic representation of a quasar observed at a certain angle $\theta$. The angle is measured starting from the accretion disc axis. If we assume the presence of an obscurer with a certain opening angle $\psi_{\rm torus}$, measured from the disc surface, it means that the inclination angle $\theta$ can vary between zero (face-on) and $\theta_{\rm max} = \pi/2 - \psi_{\rm torus}$.  }
	\label{fig: inc_sketch}
\end{figure}

To enhance the accuracy of our mock sample, we introduced in our model an obscurer, characterized by a maximum angle, $\psi_{\rm torus}$ assumed from the accretion disc, as shown in the scheme in Fig. \ref{fig: inc_sketch}. We assume this torus to be a homogeneous dust distribution that extends from the plane of the accretion disc to $\psi_{\rm torus}$. Therefore, if a quasar has an inclination angle that exceeds $\theta_{\rm max} = \pi/2 - \psi_{\rm torus}$, the torus absorbs its emission, making it undetectable\footnote{We basically assumed an infinite optical depth for the torus. This is not truly representative of the real scenario, but does not affect the consistency between our mock sample and the observed sample, as in the observed sample even mildly obscured objects are removed} 
This results in an accessible angle range of $[0,\,\theta_{\rm max})$, instead of the initial $[0,\,\pi/2)$ range. We find that by increasing $\psi_{\rm torus}$, the residual histogram becomes more and more symmetric. After some tests, we deduced that a $\theta_{\rm max}$ of $\sim$65$\degr$ (which corresponds to an obscurer angle of $\psi_{\rm torus}$=25$\degr$) is the maximum value that achieves a distribution of the residuals similar to the L20 sample, as visualized in the lower right panel of Figure \ref{fig: mock_ross_all}. This new histogram has a skewness parameter of $s = 0.19$, consistent with what we find for the L20 sample residuals histogram, which is $s_{\rm L20} = 0.20$. We note that the L20 residuals histogram is wider because while our mock sample has an overall dispersion of $\delta=0.10$ dex, L20 has $\delta=0.21$ dex. The lower left panel of Fig. \ref{fig: mock_ross_all} displays the new fit. Due to the narrower angle range, objects disperse less around the best fit, resulting in a total dispersion of $\delta_{\rm tot}=0.10$ dex. By considering this dispersion and subtracting the variability contribution quadratically, we obtain $\delta_{\rm inc}=0.06$ dex. Given the match in the residual distribution shape with the actual observed sample, we believe this to be a more reliable inclination dispersion estimate. \\
The concept of a toroidal absorber to describe the emission of quasars is not novel and it is a fundamental part of the AGN ``unified model'' (see for example \citealt{Bianchi12} for a review). At the same time, it is noteworthy that our mock sample necessitated an obscuring torus based solely on the comparison with the histogram of the residuals from the L20 sample, as this comes out as a somewhat indirect way to assess the minimum required angular width of the torus for an average population of quasars. In summary, our mock quasar sample analysis allows us to estimate the inclination contribution to the observed dispersion, approximating it at $\delta_{\rm inc} = 0.06$ dex.\\

We note that, in order to retrieve the intrinsic luminosity distribution, we could not start from the luminosity distribution of L20, as the latter is not only affected by inclination but also by additional filtering criteria that make the selection function very complex. As the latter effects cannot be corrected, we adopted instead the luminosity function of \cite{Ross13} to build our mock sample. At the same time, it is important to note that the selection of L20 does not depend on the UV luminosity itself, so it should not alter the contribution of inclination to the total observed dispersion
To address this assumption, we performed two additional tests. First, we estimated the dispersion contribution in small redshift bins ($\Delta z \sim 0.2$), in the redshift range of the L20 sample (which goes from $z\sim0$ to $z\sim5$). At all redshifts, the resulting contribution of the inclination to the final dispersion is $\delta_{\rm inc}\sim0.06$ dex, confirming the absence of a luminosity trend in this contribution. As an additional test, we build another mock sample starting from the L20 luminosity distribution, and correcting that distribution for the inclination factor. We also assumed the L20 redshift distribution to derive the redshift, and performed all the steps described above. The results are shown in the Appendix, and again a contribution of the inclination to the total dispersion of $\delta_{\rm inc}\sim0.06$ dex is confirmed. \\
The results presented here demonstrate that the dispersion due to the inclination does not depend on luminosity or redshift and that, overall, we can safely consider its contribution to be $\delta_{\rm inc}\sim0.06$ dex. Differences in the exact shape of the starting luminosity distribution for the mock sample do not seem to affect the estimate.

\section{X-ray analysis}
\label{sec:xray}

\begin{figure}
	\centering
	\includegraphics[width=.99\linewidth]{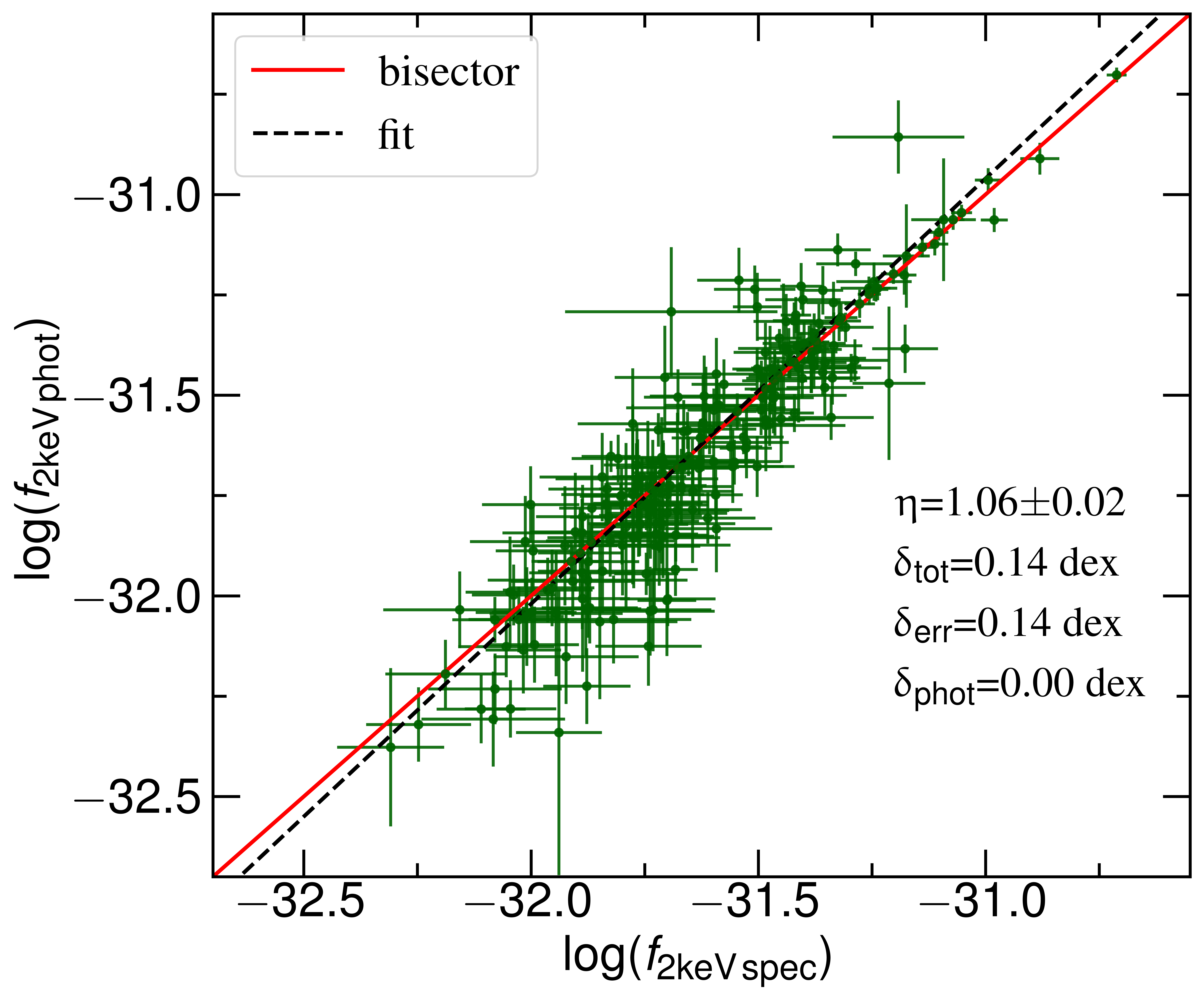}
    \caption{Comparison of the 2-keV monochromatic flux estimate derived from photometric data (see L20) and the one derived from spectroscopic data for the subsample of objects with only one observation in the 4XMM-DR9 catalogue. 
    The solid red line represents the bisector, while the black dashed line is the best fit, with a slope of $\eta=1.06\pm0.02$. The total scatter around the best fit is equal to the average uncertainties, $\delta_{\rm err} = \delta_{\rm tot} = 0.14$ dex. The ``intrinsic'' dispersion of the relation is therefore zero. 
    Overall, we can say that there is no significant contribution to the observed dispersion that comes from using photometric X-ray data instead of spectroscopic data.}
    \label{fig: Xray_singmult}
\end{figure}

Finally, another potential source of observational dispersion might arise from the use of photometric data to calculate the 2-keV monochromatic luminosities of our sources, instead of spectroscopic data. In this Section we show that this is not the case, and we can safely use photometric X-ray data. In \cite{Signorini23_qascIV}, we tested whether there is any offset between the photometric and spectroscopic flux determination. To do that, we performed the X-ray spectral analysis on a sample of 231 objects, which are all the quasars in the L20 catalogue with XMM-\textit{Newton} observations and a redshift higher than 1.9. The findings indicate that employing photometrically derived fluxes (or luminosities) instead of spectroscopically derived ones does not introduce any significant offset. However, it is possible that photometrically derived fluxes exhibit a greater scatter around the ``true value'' compared to spectroscopically derived ones. If this were the case, our prevalent use of photometric data would not introduce any bias in determining the relation parameters, but it would still increase the observed dispersion. \\
To investigate this further, we used the X-ray spectroscopically analysed sample from \cite{Signorini23_qascIV}, and we tested the equation:

\begin{equation}
\log(f_{\rm 2\,keV,\,spec}) = \eta \log(f_{\rm 2\,keV,\,phot}) + \epsilon. 
\end{equation}

We employed the same fitting procedure described in Section \ref{sec: variability}. From the 231 sources, we removed the 38 that have multiple observations. Indeed, these objects have a photometric flux estimate that has been obtained by averaging multiple observations \citep[][L20]{Lusso16}, while the spectroscopic analysis has been performed on the longest observation. As the flux estimates have been obtained in different ways, there is no point in comparing them.\\
From the fit analysis, we obtained a slope close to unity ($\eta=1.06\pm0.02$), as previously discussed in \cite{Signorini23_qascIV}.

The results are illustrated in Figure \ref{fig: Xray_singmult}. We found both $\delta_{\rm tot}$ and $\delta_{\rm err}$ to be equal to 0.14 dex, indicating that the ``intrinsic'' dispersion $\delta_{\rm phot}$ is consistent with zero. 
The outcome of this analysis is clear: when comparing the spectroscopic and photometric 2 keV flux estimates for a subsample of objects where the same observations were used to derive both estimates, there is no additional dispersion beyond that arising from observational uncertainties. Consequently, utilising photometric data instead of spectroscopic ones does not introduce any offset or bias, nor does it contribute to the total observed dispersion. This finding supports the validity and reliability of employing photometrically derived fluxes (or luminosities) in our astrophysical analyses, which is particularly relevant considering that deriving spectroscopic X-ray monochromatic flux estimates for thousands of objects is very time-consuming. \\

Another potential issue with X-ray observations is their off-axis nature. The majority of X-ray data for the L20 sample is derived from serendipitous observations, exhibiting a mean off-axis angle for the XMM-\textit{Newton} sources---which constitute most of the sample---of $\phi_{\rm offax} \sim 6.4'$. The objects not being at the detector centre might introduce additional dispersion. This was tested on a subset of 1778 objects corresponding to those with serendipitous XMM-\textit{Newton} observations in the L20 sample. We examined the $\lx$--$\luv$ relation across six off-axis angle bins: 0-2', 2'-4', 4'-6', 6'-8', 8'-10', and 10'-12'. In each bin, we observed a dispersion of $\delta \sim$ 0.24 dex, with no significant trend relating to the off-axis angle. This suggests that the off-axis angle does not substantially contribute to the dispersion. However, it is important to note that when the observed dispersion is large, minor differences become indiscernible. Assuming that objects at an off-axis angle of $\sim$10' have an additional dispersion factor of $\delta_{\rm offax}\sim0.04$ dex, this difference would be obscured in a comparison with a sample having a smaller off-axis angle, since $\sqrt{0.24^2-0.04^2}\sim0.24$, and the variance is masked by uncertainties in the dispersion estimate. One approach to further investigate this might involve analysing a subsample with a range of off-axis values but a lower initial observed dispersion, such as the 'golden sample' from \cite{Sacchi22}. However, in the 'golden sample', objects with a smaller off-axis angle (the pointed objects from \cite{nardini19}) also exhibit significantly higher average luminosity compared to those at a greater off-axis angle. Given this, along with the small overall statistics for this sample, we can not confidently distinguish the effect of lower variability due to higher average luminosity from the potential effect of dispersion reduction due to pointed observations. Future targeted observations on select subsamples might provide clearer insights into the actual impact of off-axis observations. For now, we can conclude that if this effect exists, it is likely a minimal contribution, with $\delta_{\rm offax} \leq 0.04$ dex.

\section{Comparison with observational results}
\label{sec: comparison}
We have determined that the contribution from X-ray variability to the dispersion is approximately $\delta_{\rm var} \sim 0.08$ dex. It is also evident that low-luminosity objects exhibit a greater contribution than their high-luminosity counterparts. Furthermore, our findings indicate that the use of X-ray photometric data introduces no additional dispersion to the relation. Through the construction of mock samples, we have determined that the inclination of the quasar accretion disc relative to our line of sight contributes to the total observed dispersion with $\delta_{\rm inc} \sim 0.06$ dex. The observational sources of dispersion we have assessed collectively contribute therefore with a residual dispersion of $\delta_{\rm res} = \sqrt{0.06^2+0.08^2}=0.10$ dex. We can now compare these results with the most recent estimates of the observed dispersion.

Firstly, we consider the results of \cite{Signorini23_qascIV}. In that study, using UV spectroscopic data and the best UV and X-ray proxies for the correlation, a dispersion of $\delta_{\rm obs}=0.16$ dex was found. When compared to our current estimate of the residual dispersion $\delta_{\rm res}= 0.10$ dex, it is evident that the intrinsic dispersion of the correlation must be less than $\delta_{\rm int} \leq \sqrt{\delta_{\rm obs}^{2}-\delta_{\rm res}^{2}} = 0.12$ dex.

The work of \cite{Sacchi22} offers additional clues. They presented a sample of quasars, a subsample of the L20 one, with high-quality data and a one-by-one UV and X-ray spectral analysis. This sample, which we call here the `silver sample', showed a dispersion of \(\delta_{\rm silver}=0.12\) dex. Within this sample, they also highlighted a subsample of objects at redshift \(z\sim3\), referred to as the `golden sample', with an even lower observed dispersion, \(\delta_{\rm golden} = 0.09\) dex. We display the residuals with respect to the \(\lx\)--\(\luv\) relation for these two samples in the right panels of Fig. \ref{fig: mock_ross_all}, alongside the residuals for the mock samples and for the L20 sample discussed previously. Considering the `silver' sample, if we sum up quadratically the variability (0.08 dex) and the inclination (0.06 dex), up to 0.10 dex of its dispersion can be attributed to variability and inclination. Hence, the intrinsic dispersion for the relation is estimated to be \(\delta_{\rm int} \leq \sqrt{0.12^2-0.10^2} \sim 0.07\) dex. For the golden sample, its dispersion is already smaller than the total of 0.10 dex found for our mock sample. This can be explained in terms of the sample high average luminosity, $\log(L_{\rm X}/{\rm erg\,s^{-1}\,Hz^{-1}}) \sim 27.7$. Consequently, the overall variability contribution to the dispersion is minimal, approximately \(\delta_{\rm var,H.L.} \sim 0.02\) dex, as detailed in Section \ref{sec: variability}. The intrinsic dispersion of the \(\lx\)--\(\luv\) relation for the `golden' sample can therefore be estimated as \(\delta_{\rm int} \leq \sqrt{0.09^2-0.06^2-0.02^2} \sim 0.06\) dex. Remarkably, the estimates for both subsamples are similar.

To sum up, when examining the entire quasar sample, data-quality constraints limit us to a dispersion no lower than 0.21 dex for the L20 data set and 0.16 dex for the \cite{Signorini23_qascIV} one (utilising UV spectroscopic data along with the optimal proxies for UV and X-ray emission). Of this total dispersion, 0.10 dex is attributable to the combined impact of variability and inclination. In cases of high-quality data, the dispersion can drop to 0.12 dex when the variability contribution is still significant, and to 0.09 dex when high luminosities reduce the variability contribution, as demonstrated in \cite{Sacchi22}. From these analyses, we deduce that the intrinsic dispersion of the $\lx$--$\luv$ relation must be equal to or lower than $\delta_{\rm int} \sim0.06$.

\section{Conclusions}
\label{sec: conclusions}
In this paper, we have investigated the factors that are not intrinsic to the $L_{\rm X}-L_{\rm UV}$ relation in quasars, yet might contribute to its observed dispersion and cannot be removed by selecting unbiased samples. We identified three possible dispersion sources:
\begin{itemize}
    \item Variability: quasar emission is known to be variable both in the UV and in the X-ray bands, which inevitably causes an increase in the observed dispersion. 
    Given the shorter timescales, X-ray variability is the one that is going to predominantly affect our results. To test the contribution of variability, we selected the 289 objects in the L20 sample that have multiple X-ray observations in the XMM-\textit{Newton} 4XMM-DR9 catalogue. We found that the average scatter between different estimates of the 2-keV monochromatic luminosity is $\delta_{\rm var}\sim0.08$ dex, which we can therefore consider as the variability contribution. We also found, consistently with literature results, that more luminous objects show less variability than the less luminous ones, with the ``high luminosity'' subsample having an estimate of only $\delta_{\rm var, H.L.}=0.02$ dex variability contribution to the total observed dispersion.
    \item Inclination: the inclination of the accretion disc is believed to affect the observed UV luminosity, but not the X-ray one. Therefore, the different quasar inclinations introduce a scatter in the $L_{\rm X}-L_{\rm UV}$ relation. Unfortunately, we do not have observational methods to derive inclination estimates and correct the UV luminosities accordingly. Therefore, we relied on mock-sample estimates to derive the inclination contribution to the observed dispersion.  We discussed how to recover the intrinsic luminosity distribution given an observed one, correcting for the inclination effect. Starting from the \cite{Ross13} luminosity function we then build a mock sample of quasars for which we found the inclination contribution to the dispersion to be $\delta_{\rm inc}\sim0.06$ dex, once we account for the presence of an obscuring torus. 
    \item X-ray photometry: in a previous work of this series \citep{Signorini23_qascIV}, we have shown that using photometric estimates of the 2-keV monochromatic luminosities instead of spectroscopic ones does not introduce any systematic offset (e.g. due to high offaxis angles). However, it might still introduce an additional (observed and intrinsic) dispersion. In this work we tested this for a sample of 193 objects at redshift $z>1.9$ and we found that photometric estimates do no introduce additional intrinsic additional scatter, compared to spectroscopic ones. Even though the observed scatter is larger with photometric measurements, this is accounted for by the larger uncertainties on the photometric data with respect spectroscopic ones. This is reassuring because it allows us to keep using photometric data when spectroscopic ones are not available, confident that we are not introducing additional systematic offset in the dispersion.

\end{itemize}

Comparing our results with recent observational estimates of the dispersion from \cite{Sacchi22}, we conclude that the intrinsic dispersion of the $\lx$--$\luv$ relation is exceedingly low, $\delta_{\rm int} \leq 0.06$. This finding reinforces the hypothesis that the physical mechanism governing the $\lx$--$\luv$ relation is remarkably consistent across a broad range of redshifts and luminosities. In doing such, this outcome gives additional credibility to using quasars as standard candles in cosmological studies, and it particularly highlights the significant tension existing within the flat $\Lambda$CDM cosmological model. The findings on variability confirm the results presented in \cite{Sacchi22}: subsamples with higher luminosities tend to exhibit the lowest observed dispersion values.
Looking ahead, new targeted X-ray observations of high-redshift and correspondingly high-luminosity quasars promise to yield subsamples with exceedingly low dispersion, thereby enhancing our understanding of the high-redshift Hubble diagram and associated cosmological tensions.

%
%

%

\bibliographystyle{aa} 
\bibliography{bibl}

\begin{thebibliography}{43}
\expandafter\ifx\csname natexlab\endcsname\relax\def\natexlab#1{#1}\fi

\bibitem[{{Akritas} \& {Bershady}(1996)}]{Akritas96}
{Akritas}, M.~G. \& {Bershady}, M.~A. 1996, \apj, 470, 706

\bibitem[{{Ba{\~n}ados} {et~al.}(2018){Ba{\~n}ados}, {Venemans},
  {Mazzucchelli}, {Farina}, {Walter}, {Wang}, {Decarli}, {Stern}, {Fan},
  {Davies}, {Hennawi}, {Simcoe}, {Turner}, {Rix}, {Yang}, {Kelson}, {Rudie}, \&
  {Winters}}]{Banados18}
{Ba{\~n}ados}, E., {Venemans}, B.~P., {Mazzucchelli}, C., {et~al.} 2018, \nat,
  553, 473

\bibitem[{{Bargiacchi} {et~al.}(2022){Bargiacchi}, {Benetti}, {Capozziello},
  {Lusso}, {Risaliti}, \& {Signorini}}]{bargiacchi22}
{Bargiacchi}, G., {Benetti}, M., {Capozziello}, S., {et~al.} 2022, \mnras, 515,
  1795

\bibitem[{{Bargiacchi} {et~al.}(2021){Bargiacchi}, {Risaliti}, {Benetti},
  {Capozziello}, {Lusso}, {Saccardi}, \& {Signorini}}]{bargiacchi21}
{Bargiacchi}, G., {Risaliti}, G., {Benetti}, M., {et~al.} 2021, \aap, 649, A65

\bibitem[{{Bianchi} {et~al.}(2012){Bianchi}, {Maiolino}, \&
  {Risaliti}}]{Bianchi12}
{Bianchi}, S., {Maiolino}, R., \& {Risaliti}, G. 2012, Advances in Astronomy,
  2012, 782030

\bibitem[{{Bisogni} {et~al.}(2017){Bisogni}, {Marconi}, \&
  {Risaliti}}]{Bisogni17}
{Bisogni}, S., {Marconi}, A., \& {Risaliti}, G. 2017, \mnras, 464, 385

\bibitem[{{Boyle} {et~al.}(2000){Boyle}, {Shanks}, {Croom}, {Smith}, {Miller},
  {Loaring}, \& {Heymans}}]{Boyle00}
{Boyle}, B.~J., {Shanks}, T., {Croom}, S.~M., {et~al.} 2000, \mnras, 317, 1014

\bibitem[{{Croom} {et~al.}(2009){Croom}, {Richards}, {Shanks}, {Boyle},
  {Strauss}, {Myers}, {Nichol}, {Pimbblet}, {Ross}, {Schneider}, {Sharp}, \&
  {Wake}}]{Croom09}
{Croom}, S.~M., {Richards}, G.~T., {Shanks}, T., {et~al.} 2009, \mnras, 399,
  1755

\bibitem[{{de Vries} {et~al.}(2005){de Vries}, {Becker}, {White}, \&
  {Loomis}}]{deVries05}
{de Vries}, W.~H., {Becker}, R.~H., {White}, R.~L., \& {Loomis}, C. 2005, Aj,
  129, 615

\bibitem[{{Elvis} {et~al.}(2012){Elvis}, {Hao}, {Civano}, {Brusa}, {Salvato},
  {Bongiorno}, {Capak}, {Zamorani}, {Comastri}, {Jahnke}, {Lusso}, {Mainieri},
  {Trump}, {Ho}, {Aussel}, {Cappelluti}, {Cisternas}, {Frayer}, {Gilli},
  {Hasinger}, {Huchra}, {Impey}, {Koekemoer}, {Lanzuisi}, {Le Floc'h}, {Lilly},
  {Liu}, {McCarthy}, {McCracken}, {Merloni}, {Roeser}, {Sanders}, {Sargent},
  {Scoville}, {Schinnerer}, {Schiminovich}, {Silverman}, {Taniguchi},
  {Vignali}, {Urry}, {Zamojski}, \& {Zatloukal}}]{Elvis2012}
{Elvis}, M., {Hao}, H., {Civano}, F., {et~al.} 2012, \apj, 759, 6

\bibitem[{{Foreman-Mackey} {et~al.}(2013){Foreman-Mackey}, {Hogg}, {Lang}, \&
  {Goodman}}]{emcee13}
{Foreman-Mackey}, D., {Hogg}, D.~W., {Lang}, D., \& {Goodman}, J. 2013, \pasp,
  125, 306

\bibitem[{{Giambagli} {et~al.}(2023){Giambagli}, {Fanelli}, {Risaliti}, \&
  {Signorini}}]{Giambagli23}
{Giambagli}, L., {Fanelli}, D., {Risaliti}, G., \& {Signorini}, M. 2023, arXiv
  e-prints, arXiv:2302.12582

\bibitem[{{Gianolli} {et~al.}(2023){Gianolli}, {Kim}, {Bianchi},
  {Ag{\'\i}s-Gonz{\'a}lez}, {Madejski}, {Marin}, {Marinucci}, {Matt}, {Middei},
  {Petrucci}, {Soffitta}, {Tagliacozzo}, {Tombesi}, {Ursini}, {Barnouin}, {De
  Rosa}, {Di Gesu}, {Ingram}, {Loktev}, {Panagiotou}, {Podgorny}, {Poutanen},
  {Puccetti}, {Ratheesh}, {Veledina}, {Zhang}, {Agudo}, {Antonelli},
  {Bachetti}, {Baldini}, {Baumgartner}, {Bellazzini}, {Bongiorno}, {Bonino},
  {Brez}, {Bucciantini}, {Capitanio}, {Castellano}, {Cavazzuti}, {Chen},
  {Ciprini}, {Costa}, {Del Monte}, {Di Lalla}, {Di Marco}, {Donnarumma},
  {Doroshenko}, {Dov{\v{c}}iak}, {Ehlert}, {Enoto}, {Evangelista}, {Fabiani},
  {Ferrazzoli}, {Garc{\'\i}a}, {Gunji}, {Heyl}, {Iwakiri}, {Jorstad}, {Kaaret},
  {Karas}, {Kislat}, {Kitaguchi}, {Kolodziejczak}, {Krawczynski}, {La Monaca},
  {Latronico}, {Liodakis}, {Maldera}, {Manfreda}, {Marscher}, {Marshall},
  {Massaro}, {Mitsuishi}, {Mizuno}, {Muleri}, {Negro}, {Ng}, {O'Dell},
  {Omodei}, {Oppedisano}, {Papitto}, {Pavlov}, {Peirson}, {Perri},
  {Pesce-Rollins}, {Pilia}, {Possenti}, {Ramsey}, {Rankin}, {Roberts},
  {Romani}, {Sgr{\`o}}, {Slane}, {Spandre}, {Swartz}, {Tamagawa}, {Tavecchio},
  {Taverna}, {Tawara}, {Tennant}, {Thomas}, {Trois}, {Tsygankov}, {Turolla},
  {Vink}, {Weisskopf}, {Wu}, {Xie}, \& {Zane}}]{Gianolli23}
{Gianolli}, V.~E., {Kim}, D.~E., {Bianchi}, S., {et~al.} 2023, MNRAS, 523, 4468

\bibitem[{{Gliozzi} \& {Williams}(2020)}]{Gliozzi20}
{Gliozzi}, M. \& {Williams}, J.~K. 2020, MNRAS, 491, 532

\bibitem[{{Hook} {et~al.}(1994){Hook}, {McMahon}, {Boyle}, \& {Irwin}}]{Hook94}
{Hook}, I.~M., {McMahon}, R.~G., {Boyle}, B.~J., \& {Irwin}, M.~J. 1994,
  \mnras, 268, 305

\bibitem[{{Kelly} {et~al.}(2009){Kelly}, {Bechtold}, \&
  {Siemiginowska}}]{Kelly09}
{Kelly}, B.~C., {Bechtold}, J., \& {Siemiginowska}, A. 2009, \apj, 698, 895

\bibitem[{{Lanzuisi} {et~al.}(2014){Lanzuisi}, {Ponti}, {Salvato}, {Hasinger},
  {Cappelluti}, {Bongiorno}, {Brusa}, {Lusso}, {Nandra}, {Merloni},
  {Silverman}, {Trump}, {Vignali}, {Comastri}, {Gilli}, {Schramm},
  {Steinhardt}, {Sanders}, {Kartaltepe}, {Rosario}, \&
  {Trakhtenbrot}}]{Lanzuisi14}
{Lanzuisi}, G., {Ponti}, G., {Salvato}, M., {et~al.} 2014, \apj, 781, 105

\bibitem[{{Lusso} {et~al.}(2010){Lusso}, {Comastri}, {Vignali}, {Zamorani},
  {Brusa}, {Gilli}, {Iwasawa}, {Salvato}, {Civano}, {Elvis}, {Merloni},
  {Bongiorno}, {Trump}, {Koekemoer}, {Schinnerer}, {Le Floc'h}, {Cappelluti},
  {Jahnke}, {Sargent}, {Silverman}, {Mainieri}, {Fiore}, {Bolzonella}, {Le
  F{\`e}vre}, {Garilli}, {Iovino}, {Kneib}, {Lamareille}, {Lilly}, {Mignoli},
  {Scodeggio}, \& {Vergani}}]{Lusso10}
{Lusso}, E., {Comastri}, A., {Vignali}, C., {et~al.} 2010, \aap, 512, A34

\bibitem[{{Lusso} \& {Risaliti}(2016)}]{Lusso16}
{Lusso}, E. \& {Risaliti}, G. 2016, \apj, 819, 154

\bibitem[{{Lusso} {et~al.}(2020){Lusso}, {Risaliti}, {Nardini}, {Bargiacchi},
  {Benetti}, {Bisogni}, {Capozziello}, {Civano}, {Eggleston}, {Elvis},
  {Fabbiano}, {Gilli}, {Marconi}, {Paolillo}, {Piedipalumbo}, {Salvestrini},
  {Signorini}, \& {Vignali}}]{Lusso20}
{Lusso}, E., {Risaliti}, G., {Nardini}, E., {et~al.} 2020, \aap, 642, A150

\bibitem[{{Markowitz} \& {Edelson}(2004)}]{Markowitz04}
{Markowitz}, A. \& {Edelson}, R. 2004, ApJ, 617, 939

\bibitem[{{Mortlock} {et~al.}(2011){Mortlock}, {Warren}, {Venemans}, {Patel},
  {Hewett}, {McMahon}, {Simpson}, {Theuns}, {Gonz{\'a}les-Solares}, {Adamson},
  {Dye}, {Hambly}, {Hirst}, {Irwin}, {Kuiper}, {Lawrence}, \&
  {R{\"o}ttgering}}]{Mortlock11}
{Mortlock}, D.~J., {Warren}, S.~J., {Venemans}, B.~P., {et~al.} 2011, \nat,
  474, 616

\bibitem[{{Nardini} {et~al.}(2019){Nardini}, {Lusso}, {Risaliti}, {Bisogni},
  {Civano}, {Elvis}, {Fabbiano}, {Gilli}, {Marconi}, {Salvestrini}, \&
  {Vignali}}]{nardini19}
{Nardini}, E., {Lusso}, E., {Risaliti}, G., {et~al.} 2019, \aap, 632, A109

\bibitem[{{Paolillo} {et~al.}(2017){Paolillo}, {Papadakis}, {Brandt}, {Luo},
  {Xue}, {Tozzi}, {Shemmer}, {Allevato}, {Bauer}, {Comastri}, {Gilli},
  {Koekemoer}, {Liu}, {Vignali}, {Vito}, {Yang}, {Wang}, \&
  {Zheng}}]{Paolillo17}
{Paolillo}, M., {Papadakis}, I., {Brandt}, W.~N., {et~al.} 2017, \mnras, 471,
  4398

\bibitem[{{Ponti} {et~al.}(2012){Ponti}, {Papadakis}, {Bianchi}, {Guainazzi},
  {Matt}, {Uttley}, \& {Bonilla}}]{Ponti12}
{Ponti}, G., {Papadakis}, I., {Bianchi}, S., {et~al.} 2012, \aap, 542, A83

\bibitem[{{Richards} {et~al.}(2002){Richards}, {Fan}, {Newberg}, {Strauss},
  {Vanden Berk}, {Schneider}, {Yanny}, {Boucher}, {Burles}, {Frieman}, {Gunn},
  {Hall}, {Ivezi{\'c}}, {Kent}, {Loveday}, {Lupton}, {Rockosi}, {Schlegel},
  {Stoughton}, {SubbaRao}, \& {York}}]{Richards02}
{Richards}, G.~T., {Fan}, X., {Newberg}, H.~J., {et~al.} 2002, Aj, 123, 2945

\bibitem[{{Richards} {et~al.}(2006){Richards}, {Lacy}, {Storrie-Lombardi},
  {Hall}, {Gallagher}, {Hines}, {Fan}, {Papovich}, {Vanden Berk}, {Trammell},
  {Schneider}, {Vestergaard}, {York}, {Jester}, {Anderson}, {Budav{\'a}ri}, \&
  {Szalay}}]{Richards06}
{Richards}, G.~T., {Lacy}, M., {Storrie-Lombardi}, L.~J., {et~al.} 2006, \apjs,
  166, 470

\bibitem[{{Risaliti} \& {Lusso}(2015)}]{RL15}
{Risaliti}, G. \& {Lusso}, E. 2015, \apj, 815, 33

\bibitem[{{Risaliti} \& {Lusso}(2019)}]{RL19_nature}
{Risaliti}, G. \& {Lusso}, E. 2019, Nature Astronomy, 3, 272

\bibitem[{{Ross} {et~al.}(2013){Ross}, {McGreer}, {White}, {Richards}, {Myers},
  {Palanque-Delabrouille}, {Strauss}, {Anderson}, {Shen}, {Brandt},
  {Y{\`e}che}, {Swanson}, {Aubourg}, {Bailey}, {Bizyaev}, {Bovy}, {Brewington},
  {Brinkmann}, {DeGraf}, {Di Matteo}, {Ebelke}, {Fan}, {Ge}, {Malanushenko},
  {Malanushenko}, {Mandelbaum}, {Maraston}, {Muna}, {Oravetz}, {Pan},
  {P{\^a}ris}, {Petitjean}, {Schawinski}, {Schlegel}, {Schneider}, {Silverman},
  {Simmons}, {Snedden}, {Streblyanska}, {Suzuki}, {Weinberg}, \&
  {York}}]{Ross13}
{Ross}, N.~P., {McGreer}, I.~D., {White}, M., {et~al.} 2013, \apj, 773, 14

\bibitem[{{Sacchi} {et~al.}(2022){Sacchi}, {Risaliti}, {Signorini}, {Lusso},
  {Nardini}, {Bargiacchi}, {Bisogni}, {Civano}, {Elvis}, {Fabbiano}, {Gilli},
  {Trefoloni}, \& {Vignali}}]{Sacchi22}
{Sacchi}, A., {Risaliti}, G., {Signorini}, M., {et~al.} 2022, \aap, 663, L7

\bibitem[{{Sanders} {et~al.}(1989){Sanders}, {Phinney}, {Neugebauer}, {Soifer},
  \& {Matthews}}]{Sanders89}
{Sanders}, D.~B., {Phinney}, E.~S., {Neugebauer}, G., {Soifer}, B.~T., \&
  {Matthews}, K. 1989, \apj, 347, 29

\bibitem[{{Shakura} \& {Sunyaev}(1973)}]{Shakura73}
{Shakura}, N.~I. \& {Sunyaev}, R.~A. 1973, \aap, 24, 337

\bibitem[{{Signorini} {et~al.}(2023){Signorini}, {Risaliti}, {Lusso},
  {Nardini}, {Bargiacchi}, {Sacchi}, \& {Trefoloni}}]{Signorini23_qascIV}
{Signorini}, M., {Risaliti}, G., {Lusso}, E., {et~al.} 2023, arXiv e-prints,
  arXiv:2306.16438

\bibitem[{{Sobolewska} \& {Done}(2007)}]{Sobolewska07}
{Sobolewska}, M.~A. \& {Done}, C. 2007, MNRAS, 374, 150

\bibitem[{{Steffen} {et~al.}(2006){Steffen}, {Strateva}, {Brandt}, {Alexander},
  {Koekemoer}, {Lehmer}, {Schneider}, \& {Vignali}}]{Steffen06}
{Steffen}, A.~T., {Strateva}, I., {Brandt}, W.~N., {et~al.} 2006, \aj, 131,
  2826

\bibitem[{{Tananbaum} {et~al.}(1979){Tananbaum}, {Avni}, {Branduardi}, {Elvis},
  {Fabbiano}, {Feigelson}, {Giacconi}, {Henry}, {Pye}, {Soltan}, \&
  {Zamorani}}]{Tananbaum79}
{Tananbaum}, H., {Avni}, Y., {Branduardi}, G., {et~al.} 1979, \apjl, 234, L9

\bibitem[{{Vagnetti} {et~al.}(2011){Vagnetti}, {Turriziani}, \&
  {Trevese}}]{Vagnetti11}
{Vagnetti}, F., {Turriziani}, S., \& {Trevese}, D. 2011, \aap, 536, A84

\bibitem[{{Vanden Berk} {et~al.}(2004){Vanden Berk}, {Wilhite}, {Kron},
  {Ivezic}, {Pereyra}, \& {SDSS Collaboration}}]{VandenBerk04}
{Vanden Berk}, D., {Wilhite}, B., {Kron}, R., {et~al.} 2004, in American
  Astronomical Society Meeting Abstracts, Vol. 205, American Astronomical
  Society Meeting Abstracts, 120.02

\bibitem[{{Wang} {et~al.}(2021){Wang}, {Yang}, {Fan}, {Hennawi}, {Barth},
  {Banados}, {Bian}, {Boutsia}, {Connor}, {Davies}, {Decarli}, {Eilers},
  {Farina}, {Green}, {Jiang}, {Li}, {Mazzucchelli}, {Nanni}, {Schindler},
  {Venemans}, {Walter}, {Wu}, \& {Yue}}]{Wang21}
{Wang}, F., {Yang}, J., {Fan}, X., {et~al.} 2021, \apjl, 907, L1

\bibitem[{{Webb} {et~al.}(2020){Webb}, {Coriat}, {Traulsen}, {Ballet}, {Motch},
  {Carrera}, {Koliopanos}, {Authier}, {de la Calle}, {Ceballos}, {Colomo},
  {Chuard}, {Freyberg}, {Garcia}, {Kolehmainen}, {Lamer}, {Lin}, {Maggi},
  {Michel}, {Page}, {Page}, {Perea-Calderon}, {Pineau}, {Rodriguez}, {Rosen},
  {Santos Lleo}, {Saxton}, {Schwope}, {Tom{\'a}s}, {Watson}, \&
  {Zakardjian}}]{Webb20}
{Webb}, N.~A., {Coriat}, M., {Traulsen}, I., {et~al.} 2020, \aap, 641, A136

\bibitem[{{Young} {et~al.}(2009){Young}, {Elvis}, \& {Risaliti}}]{Young09}
{Young}, M., {Elvis}, M., \& {Risaliti}, G. 2009, Apj, 183, 17

\bibitem[{{Young} {et~al.}(2010){Young}, {Elvis}, \& {Risaliti}}]{Young2010}
{Young}, M., {Elvis}, M., \& {Risaliti}, G. 2010, \apj, 708, 1388

\end{thebibliography}

\appendix
\section{Correction of the  Luminosity Function for inclination effects} 
\label{appendix: LF}
In this Appendix we report how we derive the ``intrinsic'' luminosity function $m(\mathcal{L})$ from the observed one, $n(L)$. We recall that the observed luminosity function consists of objects where the inclination effect has modified each object observed luminosity, while we are interested in recovering the intrinsic luminosity distribution.
We start from the relation between $n(L)$ and  $m(\mathcal{L})$ . 
\begin{equation}
	n(L)dL = \int_{0}^{\pi/2} m\left(\frac{L}{\cos\theta}\right)\sin\theta \,d\theta d\mathcal{L}
	\label{f2}
\end{equation}

We substitute $\cos\theta = x$ and $dx = -\sin\theta\,d\theta$:

\begin{equation}
n(L) = \int_{0}^{1} m(L/x) dx
\label{f2}
\end{equation}

We can now write that $L/x = \mathcal{L}$ and $-(L/x^2)$ d$x =$ d$\mathcal{L}$ so that

\begin{equation}
    n(L) = \int_{L}^{\infty} m(\mathcal{L}) \frac{L}{\mathcal{L}^2} d\mathcal{L}
\label{f3}
\end{equation}

We derive with respect to $L$:

\begin{equation}
    \frac{\partial}{\partial L} n(L) = \frac{\partial}{\partial L} \int_{L}^{\infty} m(\mathcal{L}) \frac{L}{\mathcal{L}^2} d\mathcal{L} = \int_{L}^{\infty} \frac{m(\mathcal{L})}{\mathcal{L}^2} d\mathcal{L} - \frac{m(L)}{L}
\label{f4}
\end{equation}

and then derive again:

\begin{equation}
    \frac{\partial^2}{\partial L^2} n(L) = -\frac{m(L)}{L^2} - \frac{L m'(L)-m(L)}{L^2} = -\frac{m'(L)}{L}
\label{f5}
\end{equation}

from which:
\begin{equation}
    m(\mathcal{L}) =  \int_{L}^{\infty} n''(\mathcal{L}) \mathcal{L} d\mathcal{L}
\label{f6}    
\end{equation}

and finally get:
\begin{equation}
m(\mathcal{L}) = n(L) - n'(L) L 
\label{FM}
\end{equation}

\section{Additional material} \label{appendixB}
\begin{figure*}
	\centering
	\subfloat{{\includegraphics[width=8.75cm]{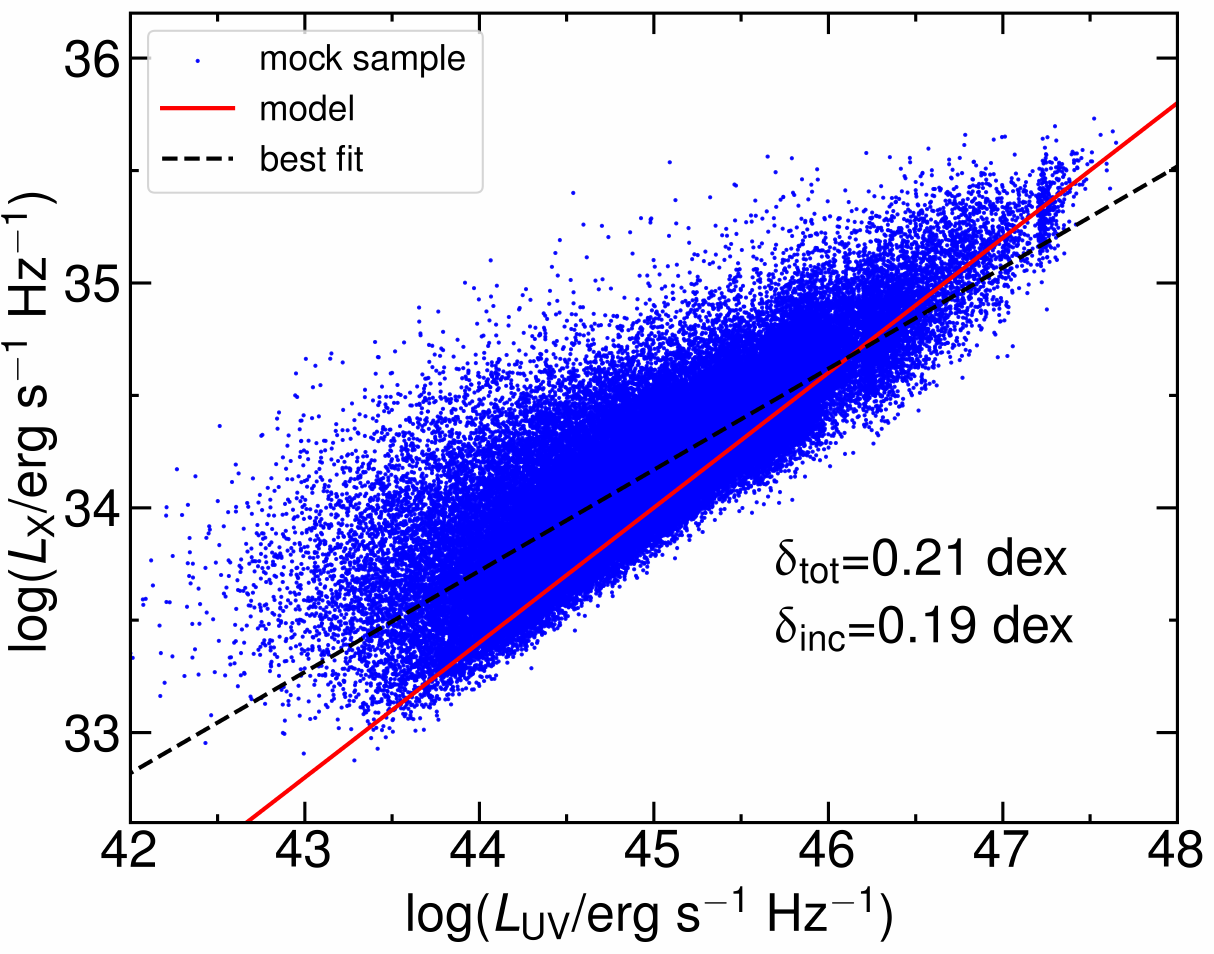} }}%
	\qquad
	\subfloat{{\includegraphics[width=8.75cm]{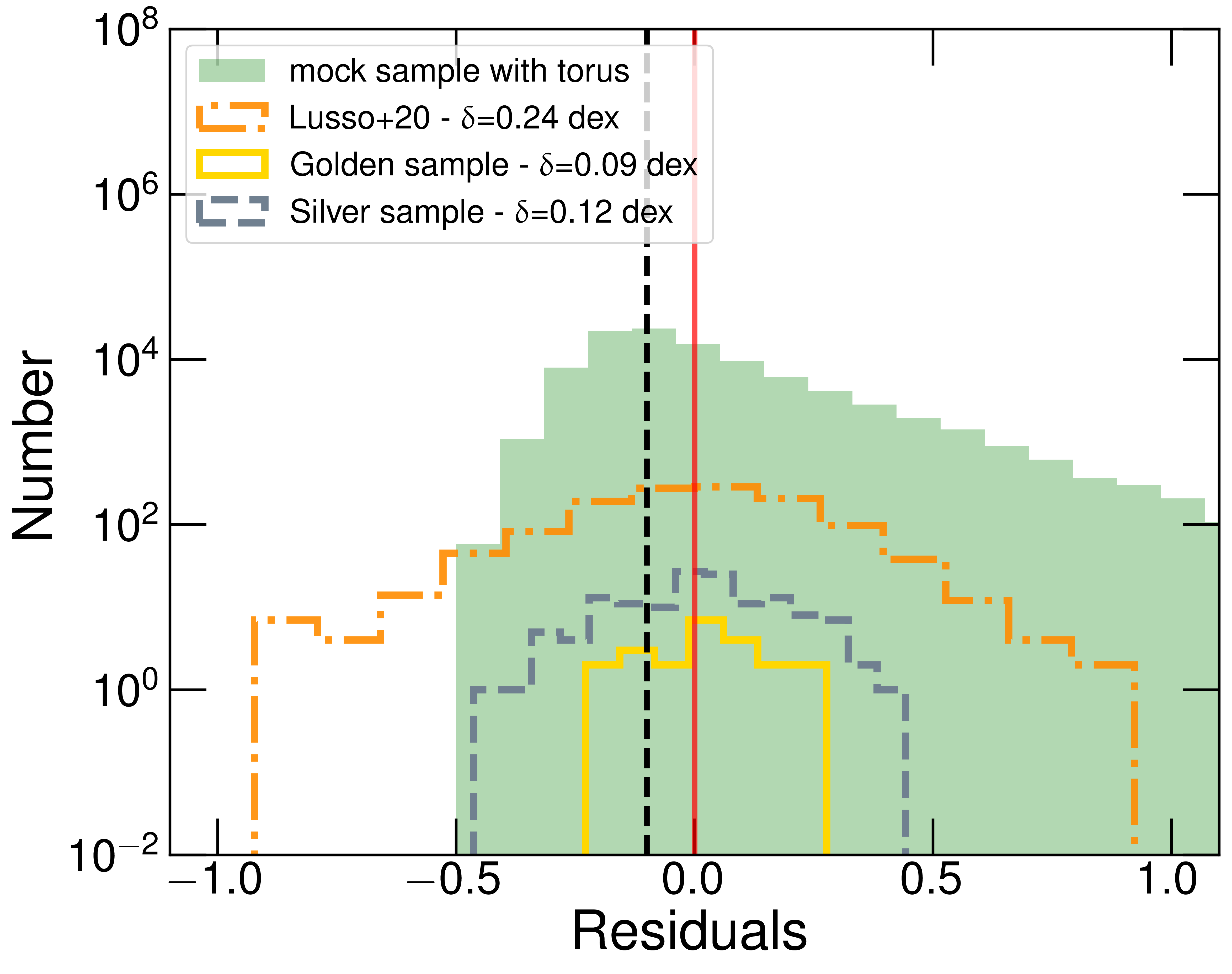} }}%
	\qquad
	\subfloat{{\includegraphics[width=8.75cm]{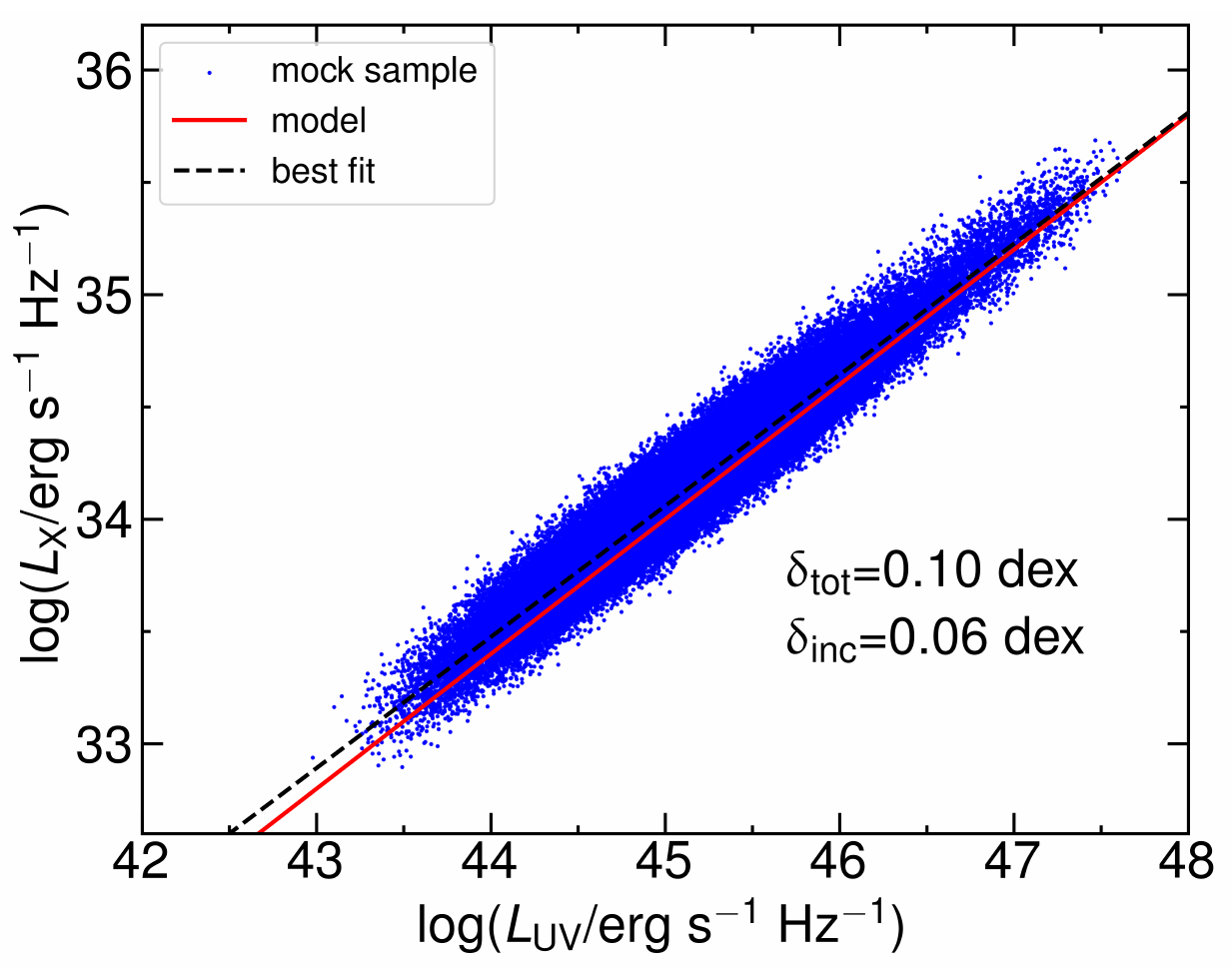} }}%
	\qquad
	\subfloat{{\includegraphics[width=8.75cm]{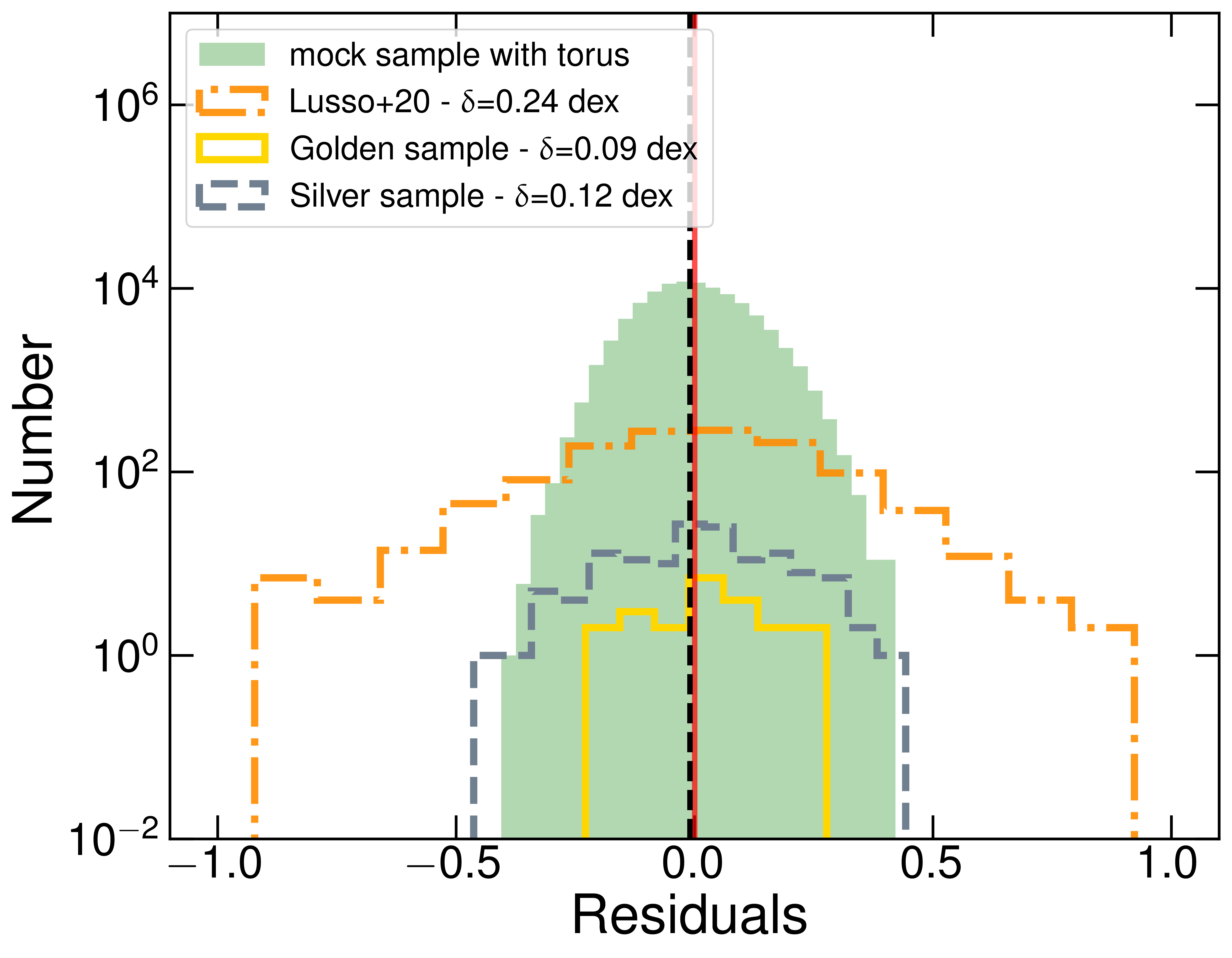} }}%
	\caption{\textit{Upper left panel}: mock sample of 100.000 quasars assuming an intrinsic relation with zero intrinsic dispersion, a contribution from variability to the observed dispersion of 0.08 dex, and an inclination angle between 0 and 90°. For each object, a redshift is assigned using the L20 sample redshift distribution. To derive the luminosity, we correct the L20 luminosity distribution for th einclination effect, and extract a luminosity value from there. The red solid line represents the starting sample, with a slope $\gamma=0.6$ and a zero dispersion. The blue points show the sample after the dispersion due to variability is added and the objects are assigned a random inclination. The total dispersion of the sample is $\delta_{\rm tot}=0.21$ dex, and the inclination effect accounts for $\delta_{\rm inc}$=0.19 dex. \\
	\textit{Upper right panel}: in green filled, the histogram of the fit residuals. This distribution is skewed, with a skewness parameter of $s=1.74$, and the peak is shifter from zero, at -0.16. In dot-dashed orange, the residuals histogram for the L20 sample. We see that the L20 distribution, which is the observed one, is instead much more symmetric, with a skewness parameter of $s_{\rm L20} = 0.20$. In dashed silver and in solid gold, the residuals histograms corresponding to the "silver sample" and the "golden sample" of \cite{Sacchi22} (details in Section \ref{sec: comparison}). The red solid vertical line corresponds to zero, while the dashed black vertical line corresponds to the peak of the mock sample distribution, equal to -0.1.  \\  
	\textit{Lower left panel}: the same as the Upper Left panel, but assuming the presence of an obscurer with an angle width of $\psi_{\rm torus}=25\degr$, which means that the inclination angle for the objects in the mock sample can go from 0 to 65$\degr$.  The total dispersion is reduced to $\delta_{\rm tot}=0.10$ dex, and the inclination effect accounts for $\delta_{\rm inc}$=0.06 dex. \\
	\textit{Lower right panel}: the same as the Upper right panel, but assuming the presence of an obscurer with an angle width of $\psi_{\rm torus}=25\degr$. Here we see that the residuals distribution is symmetric, with a skewness parameter of $s=0.20$, which perfectly matches the L20 value of $s_{\rm L20} = 0.20$. We also note that the residuals distribution is now slightly wider than the one for the \cite{Sacchi22} sample.
 \label{fig: appendix_lusso}
}
	\label{fig: mock_lusso}%
\end{figure*}

We provide here the fit of the $\lx$--$\luv$ relation and the histograms of the residuals for the mock samples derived by starting from L20 luminosity distribution. In Fig. \ref{fig: appendix_lusso} we show the results obtained without the torus assumption in the upper panel, and with the assumption of a torus with an angle width of $\psi_{\rm torus} = 25\degr$ in the lower panels. The histograms scale is set to logarithmic to better visualise the different shapes. In the Upper left panel, we see that without the torus assumption, we obtain a high estimate of the dispersion due to the inclination, $\delta_{\rm inc}=0.19$ dex, and a slope of the relation equal to $\gamma = 0.46 \pm 0.01$, not consistent with the assumed value of $\gamma = 0.6$ for the mock sample. In the Upper right panel we also see that, as in the case discussed in the text where we started from the \cite{Ross13} luminosity function, without the torus assumption we obtain a highly skewed histograms of the residuals, with a skewness parameter of $s = 1.74$. The peak of the histogram is also shifted from zero, and is found at $-0.1$.  \\
In the lower panel, we show the results obtained assuming a $\psi_{\rm torus} = 25\degr$ torus. In the lower left panel, we see that the best fit slope is now $\gamma = 0.60\pm0.01$, perfectly consistent with the assumed value of $\gamma=0.6$. We note that with this assumption, the dispersion due to inclination is estimated to be $\delta_{\rm inc}= 0.06$ dex, which is the exact same value that we obtain starting from \cite{Ross13} luminosity distribution. In the lower right panel, we see now that the histograms of the residuals is now symmetric, with a skewness parameter of $s = 0.20$ and with the peak corresponding to $-0.06$. \\
To sum up, starting with the L20 luminosity distribution instead of with the \cite{Ross13} luminosity function, we obtain the same results both in terms of the need for a $\psi_{\rm torus} \sim 25 \degr$ torus, and for the estimate of the contribution of inclination to the total dispersion, with $\delta_{\rm inc} = 0.06$. This result shows us that our estimate for the inclination contribution does not strongly depend on the exact shape of the starting luminosity distribution.

%

\end{document}